\documentclass[12pt,preprint]{aastex}
\usepackage{epsfig}
\usepackage{natbib}
\usepackage{graphicx}
\usepackage{slashbox}
\usepackage{multirow}
\usepackage{lscape}
\usepackage{mathrsfs,amssymb}
\usepackage{amsmath}
\newcommand       \Angstrom     {\,{\rm \AA}}

\newcommand       \cm           {\,{\rm cm}}

\newcommand       \km           {\,{\rm km}}

\newcommand       \K            {\,{\rm K}}

\newcommand       \kpc          {\,{\rm kpc}}
\newcommand       \s            {\,{\rm s}}

\newcommand       \yr       {\,{\rm yr}}

\newcommand       \gtsim        {\gtrsim}

\newcommand       \mum          {\,{\rm \mu m}}

\newcommand       \Teff         {T_{\rm eff}}
\newcommand       \msun         {\,{M_\odot}}

\newcommand       \simali       {\sim\,}

\def    \W       {\,{\rm W}}
\def    \m       {\,{\rm m}}

\def    \Md      {M_{\rm dust}}
\def    \Mdloss  {\dot{M}_{\rm dust}}

\def    \Mloss   {\dot{M}}
\def    \dMAGB   {\dot{M}_{\scriptscriptstyle\rm AGB}}
\def    \dMSW    {\dot{M}_{\scriptscriptstyle\rm SW}}
\def    \tauAGB  {\tau_{\scriptscriptstyle\rm AGB}}
\def    \tauSW   {\tau_{\scriptscriptstyle\rm SW}}
\def    \MdustAGB  {{M}_{\rm dust}^{\scriptscriptstyle\rm AGB}}
\def    \MdustSW  {{M}_{\rm dust}^{\scriptscriptstyle\rm SW}}
\def    \gastodust  {M_{\rm gas}/M_{\rm dust}}
\def    \cotoHH  {n({\rm CO})/n({\rm H_2})}
\def    \cctoHH  {n({\rm C_2})/n({\rm H_2})}
\def    \cntoHH  {n({\rm CN})/n({\rm H_2})}
\def    \rhoco   {\rho_{\rm CO}}
\def    \Ftwenty     {F_{\scriptscriptstyle 21}}
\def    \Fthirty     {F_{\scriptscriptstyle 30}}
\def    \Fuir     {F_{\scriptscriptstyle\rm UIR}}
\def    \Fir      {F_{\scriptscriptstyle\rm IR}}

\def    \rmin           {r_{\rm min}}
\def    \rmax           {r_{\rm max}}
\def    \rsw            {r_{\rm sw}}
\def    \vexp           {v_{\rm exp}}
\countdef\decade=200
\advance\decade by \year
\countdef\hours=201
\advance\hours by \time
\divide\hours by 60
\countdef\mins=202
\advance\mins by \hours
\multiply\mins by 60
\multiply\hours by 100
\countdef\miltime=203
\advance\miltime by \hours
\advance\miltime by \time
\advance\miltime by -\mins
\def\today{\number\decade.\number\month.\number\day.\number\miltime}

\shorttitle{On the Relation of Mysterious 21$\mum$ Features}
\title{
\vspace*{-2.0em}
{\normalsize\rm Accepted for publication in
               {\it The Astrophysical Journal}}\\
\vspace*{1.0em}
On the Relation between the Mysterious 21$\mum$
       Emission Feature of Post-Asymptotic Giant Branch Stars
       and Their Mass Loss Rates
\\{\small DRAFT: \today ~~}
}
\author{Ajay Mishra\altaffilmark{1},
        Aigen Li\altaffilmark{1},
        and B.W.~Jiang\altaffilmark{2}}
\altaffiltext{1} {Department of Physics and Astronomy,
                  University of Missouri,
                  Columbia, MO 65211, USA;
                  {\sf amishra@mail.missouri.edu,
                       lia@missouri.edu}
                   }
\altaffiltext{2}{Department of Astronomy,
                 Beijing Normal University,
                 Beijing 100875, China;
                 {\sf bjiang@bnu.edu.cn}
                 }

\begin{document}

\begin{abstract}
%
Over two decades ago, a prominent, mysterious emission band
peaking at $\simali$20.1$\mum$ was serendipitously
detected in four preplanetary nebulae 
(PPNe; also known as ``protoplanetary nebulae''). 
%
To date, this spectral feature,
designated as the ``21$\mum$'' feature, 
has been seen in 27 carbon-rich PPNe
in the Milky Way and the Magellanic Clouds.
The nature of its carriers
remains unknown although many candidate materials 
have been proposed.
The 21$\mum$ sources also exhibit
an equally mysterious, unidentified emission feature
peaking at 30$\mum$. 
%
While the 21$\mum$ feature is {\it exclusively} 
seen in PPNe, a short-lived evolutionary stage 
between the end of the asymptotic giant branch (AGB)
and planetary nebula (PN) phases,
the 30$\mum$ feature is more commonly observed
in all stages of stellar evolution from 
the AGB through PPN to PN phases.
%
%
%
We derive the stellar mass loss rates ($\Mloss$)
of these sources from their infrared (IR) emission,  
using the ``2-DUST'' radiative transfer code 
for axisymmetric dusty systems
which allows one to distinguish 
the mass loss rates
of the AGB phase ($\dMAGB$)
from that of the superwind ($\dMSW$) phase.
We examine the correlation 
between $\dMAGB$ or $\dMSW$
and the fluxes emitted from  
the 21 and 30$\mum$ features.
We find that both features tend to
correlate with $\dMAGB$, 
suggesting that their carriers 
are probably formed in the AGB phase. 
%
The nondetection of the 21$\mum$ feature
in AGB stars suggests that, unlike the 30$\mum$ feature,
the excitation of the carriers of 
the 21$\mum$ feature
may require ultraviolet photons 
which are available in PPNe but not in AGB stars. 
\end{abstract}


\section{Introduction\label{sec:intro}}
%
%
%
Historically, the evolutionary stage
of low- to intermediate-mass stars ($\simali$0.8--8$\msun$)
in between the end of the asymptotic giant branch 
(AGB) phase and the planetary nebula (PN) phase
has been designated as 
the proto-planetary nebula phase.
This short evolutionary phase
of $\simali$$10^{3}\yr$ is now also called
``preplanetary nebula''
(e.g., see Sahai et al.\ 2005)
because the phrase ``proto-planetary'' 
is widely used by the exoplanetary and 
planet formation communities to refer to 
dusty disks around young stars.
Following Sahai et al.\ (2005),
in this work we will use the term 
``preplanetary nebula''  (PPN).
We also note that, 
in the literature, the term preplanetary 
(or proto-planetary) nebulae (PPNe) has been 
interchangeably used with the term post-AGB objects.
%
%
%
%

The so-called ``21$\mum$'' PPNe, 
a class of C-rich PPNe
which all exhibit a prominent emission feature 
at 21$\mum$, have received increasing attention 
over the past decades. 
%
This feature was serendipitously discovered 
by Kwok et al.\ (1989) in four PPNe 
through the 7.7--22.6$\mum$ spectra 
obtained by the {\it Low Resolution Spectrometer} (LRS)
on board the {\it Infrared Astronomical Satellite} (IRAS).
So far, this feature has been seen unambiguously 
in 27 carbon-rich PPNe, 
including 18 Galactic objects (Cerrigone et al.\ 2011)
and nine objects in the Large and Small
Magellanic Clouds (Volk et al.\ 2011).
The spectral profile 
of the so-called ``21$\mum$'' feature 
displays little variation
among different sources:
it always peaks at $\simali$20.1$\mum$ 
and has a more or less constant FWHM 
(full width at half maximum)   
of $\simali$2.2--2.3$\mum$. 
%
%

The exact nature of the carriers 
of this feature remains unknown
ever since its discovery in 1989.
This feature is potentially an important
probe of the physical and chemical processes
occurring in PPNe, a short evolutionary stage
immediately preceding the PN phase.
Also, it is important in terms of energetics
since this feature emits up to $\sim$\,8\% of 
the total infrared (IR) power of a 21$\mum$ source.
The fact that so much power is carried in this feature
suggests that its carrier must be made of 
abundant elements (e.g., C).
%
%
%
Over the past decades,
over a dozen carrier
candidates have been suggested
(e.g., see Posch et al.\ 2004, 
Zhang et al.\ 2009a).
However, none of them can be considered viable
because they either could not reproduce 
the observed spectral profile 
(e.g., SiC [see Jiang et al.\ 2005],
FeO [see Li et al.\ 2013])
or they require too much dust material 
(e.g., TiC [see Li 2003]).
%
%

A unique characteristics of 
the mysterious 21$\mum$ feature is that 
it is so far only detected in PPNe,
neither in the AGB phase nor in the PN phase.
This might suggest that the condensation of
the carriers of the 21$\mum$ feature
may occur during the so-called superwind phase
(e.g., see von Helden et al.\ 2000), 
a phase of high mass-loss where AGB stars 
lose the remaining stellar envelope
and terminate their life on the AGB 
(see Renzini 1981).
It is also possible that,
after entering the PN phase, 
the 21$\mum$-feature carriers
may be rapidly destroyed 
by the highly energetic photons available in PNe.
%
%
%
To examine this hypothesis, in this work
we derive the stellar mass loss rates of 
the 21$\mum$ PPNe during the AGB phase 
as well as the superwind phase
and explore the correlation 
between the 21$\mum$ feature
and the mass loss rates,
with an aim of gaining insight into 
the formation and destruction of 
the carriers of this mysterious feature.

%
%
%

\section{The 21$\mum$ Sources: 
         Photometry and Spectroscopy}
         \label{sec:sample}
We will derive the mass loss rates of 
the 21$\mum$ sources from their infrared (IR)
emission spectral energy distributions (SEDs). 
In this work we will consider all the 18 Galactic 
21$\mum$ sources. The IR emission of these sources 
have been measured with various instruments
through broadband photometry: 
(i) the {\it Infrared Astronomical Satellite} (IRAS) 
at 12, 25, 60, and 100$\mum$,
(ii) the {\it Midcourse Space Experiment} (MSX)
$A$, $C$, $D$, and $E$ bands respectively
at 8.28, 12.13, 14.65 and 21.3$\mum$,
(iii) the {\it Infrared Camera} (IRC) 
on board the AKARI satellite at 9 and 18$\mum$, and 
(iv) the {\it Wide-field Infrared Survey Explorer} 
(WISE) at 3.4, 4.6, 12 and 22$\mum$. 
Six of these 18 sources have also been measured 
at 1.2\,mm with the {\it Max-Planck Millimeter Bolometer} 
(MAMBO) array at the 30-meter IRAM telescope
(Buemi et al.\ 2007).
We compile from the literature
the stellar and circumstellar parameters
of all the 18 Galactic 21$\mum$ sources
and tabulate them in Table~\ref{tab:starpara}. 
These include the stellar effective temperature $\Teff$, 
stellar luminosity $L_\star$, 
stellar core mass $M_\star$,
stellar radius $r_\star$, 
and distance $d$ from Earth 
of the central star.
Also tabulated in Table~\ref{tab:starpara}
are the {\it IRAS}, {\it MSX}, {\it AKARI}, 
{\it WISE} and {\it IRAM}/MAMBO photometry.

In addition to the 21$\mum$ feature,
the 21$\mum$ sources also display a prominent, 
mysterious emission feature at 30$\mum$.
The 30$\mum$ feature
is very broad and strong
and extends from $\simali$24$\mum$ 
to $\simali$45$\mum$.
It often accounts for up to $\simali$30\% of 
the total IR luminosity 
of such an object (Volk et al.\ 2002).
%
%

The 21$\mum$ sources also exhibit a distinctive set of 
emission features at 3.3, 6.2, 7.7, 8.6, and 11.3$\mum$
(Hrivnak et al.\ 2008, Volk 2015).
Theses features are collectively known as
the ``unidentified infrared'' (UIR) features
and commonly attributed to 
polycyclic aromatic hydrocarbon (PAH) molecules 
(L\'eger \& Puget 1984, 
Allamandola et al.\ 1985). 

For these sources, high-quality mid-IR spectra 
have obtained with the {\it Infrared Spectrograph} 
(IRS) on board the {\it Spitzer Space Telescope} 
and the {\it Short Wavelength Spectrometer} (SWS)
on board the {\it Infrared Space Observatories} (ISO).
These spectra allow one to measure relatively accurately 
the (integrated) fluxes emitted 
in the 21$\mum$ feature ($\Ftwenty$)
and the 30$\mum$ feature ($\Fthirty$),
and to a less degree, 
the UIR features ($\Fuir$).\footnote{%
   For some sources the {\it Spitzer}/IRS 
   or {\it ISO}/SWS data are available only
   at $\lambda\gtsim10\mum$. Therefore,
   for these sources $\Fuir$ is underestimated 
   as the UIR bands at 6.2, 7.7 and 8.6$\mum$
   are not counted in $\Fuir$.
   }

\begin{figure*}
\centerline{
\includegraphics[width=17.0cm]{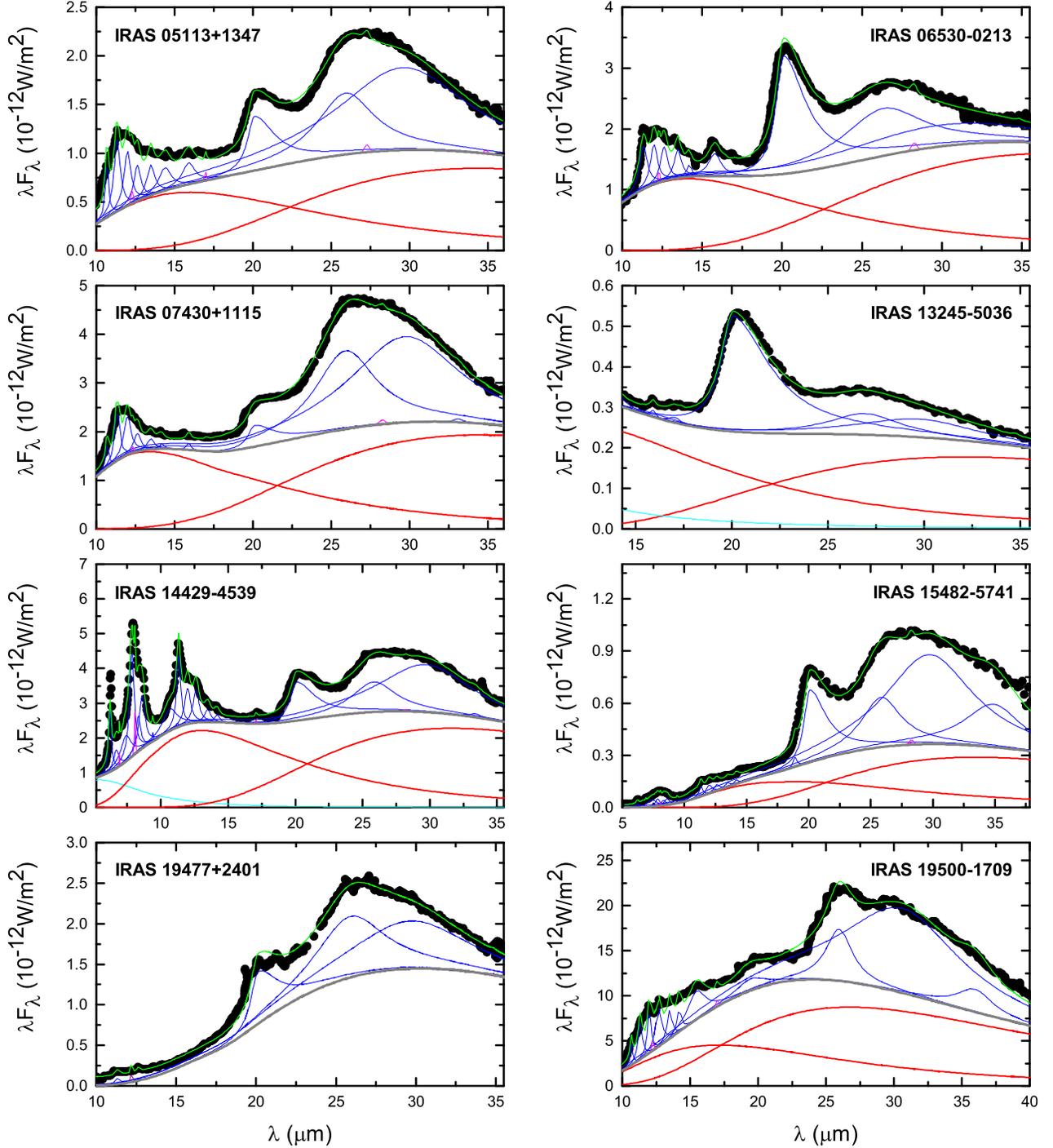}
}
\caption{\footnotesize
         \label{fig:drudefit}
        Decomposition of the observed IR spectra (black points)
	of eight 21$\mum$ sources 
        (IRAS 05113+1347, IRAS 06530-0213,
	IRAS 07430+1115, IRAS 13245-5036, IRAS 14429-4539, 
	IRAS 15482-5741, IRAS 19477+2401, and IRAS 19500-1709) 
	into a stellar continuum (cyan line), two dust thermal 
	continuum emission components (red lines), the 21, 
	30$\mum$ and UIR features (blue lines), and the H$_{2}$ 
	bands (magenta lines). 
        The decomposition is done with 
	the PAHFIT technique. Gray lines are the sum of the 
	stellar and dust thermal continuum. Green lines show 
	the resulting model spectra.
        For most sources (except IRAS 13245-5036
        and IRAS 14429-4539), the stellar contribution is 
        too weak to show up. For IRAS 19477+2401, there is
        only one dust continuum component 
        and the stellar continuum is so small that the gray line 
        (which plots the sum of the stellar and dust continuum) 
        overlaps the red line (which plots the dust continuum).  
        }
\end{figure*}

Mishra, Li \& Jiang (2015) had used 
the PAHFIT software of Smith et al.\ (2007) 
to decompose 
the {\it Spitzer}/IRS or {\it ISO}/SWS spectra
of ten Galactic 21$\mum$ sources and had already 
determined $\Ftwenty$, $\Fthirty$, and $\Fuir$
for these ten sources.
Following Mishra, Li \& Jiang (2015),
we analyze the {\it Spitzer}/IRS or 
{\it ISO}/SWS spectra of the remaining 
eight sources and decompose their IR spectra into
(i) a stellar continuum $B_\lambda(\Teff)$ 
    which is approximated by a blackbody  
    at the stellar effective temperature $\Teff$,
(ii) a thermal continuum of warm dust 
     of temperature $T_W$  
     represented by a modified blackbody
     $\lambda^{-2}\,B_\lambda(T_W)$,
(iii) a thermal continuum of cold dust 
      of temperature $T_C$
      represented by 
      $\lambda^{-2}\,B_\lambda(T_C)$, and 
(iv) the 21$\mum$, 30$\mum$ and UIR features
     approximated by a set of Drude functions.
In addition, the H$_2$ S(0)--S(7) rotational lines
are included and approximated by a number of 
Gaussian files. 
In Figure~\ref{fig:drudefit} 
we show the spectral decompositional fits 
to the eight sources.
In Table~\ref{tab:drudefit}
we tabulate the fluxes emitted 
in the 21$\mum$, 30$\mum$ and UIR features. 

\section{Stellar Mass Loss Rates}\label{sec:massloss}
We employ the ``2-DUST'' radiative transfer code 
to model the dust IR emission of 
the selected 21$\mum$ sources,
using the {\it IRAS} 
(and the 1.2\,mm IRAM/MAMBO, if available) 
photometric data.
The ``2-DUST'' code, 
developed by Ueta \& Meixner (2003) 
for dusty axisymmetric systems,
is well suited for modeling the IR emission of
the PPN layered dust shells. 

Following Ueta \& Meixner (2003), we consider 
a dust density function that has 
(i) a spherical outer shell 
-- the remnant of the AGB wind, 
(ii) a spheroidal mid-region, and 
(iii) an inner toroidal core
created during the superwind phase
-- a rather brief period of 
equatorially-enhanced mass-loss 
near the end of the AGB mass-loss phase.
The mid-region of the shell assumes
a somewhat spheroidal distribution,
reflecting the transition 
from a spherical mass-loss geometry
to an axial symmetry during 
the course of the AGB mass-loss history.
The underlying assumption of
this axisymmetric density distribution function 
is that the mass loss 
in the AGB phase is spherically symmetric 
and converted to axisymmetric after the end 
of the AGB phase (i.e., the ``superwind'' phase). 
The adopted density distribution is a function
of the radius of the dust shell $r$,
the latitudinal angle $\Theta$, 
and five geometric parameters
($A$, $B$, $C$, $D$, and $E$;
see Meixner et al.\ 2002,
Ueta \& Meixner 2003): 
%
%

\begin{equation}
\label{eq:dndr}
\begin{split}
\rho(r,\Theta) = \rho_{\rm min}\left(\frac{r}{r_{\rm min}}\right)^{-B\left\{1+C\sin^{F}\Theta\left[e^{-\left(r/\rsw\right)^{D}}/e^{-\left(\rmin/\rsw\right)^{D}}\right]\right\}}\\
\times \left\{ 1+A\left(1-\cos\Theta\right)^F\left[e^{-\left(r/\rsw\right)^{E}}/e^{-\left(\rmin/\rsw\right)^{E}} \right] \right\} ~,
\end{split}
\end{equation}
where $\rho(r,\Theta)$ is the dust mass density 
at radius $r$ and latitude $\Theta$, 
$\rho_{\rm min}$ is the dust mass density on 
the polar axis at the inner edge of the shell, 
$r_{\rm min}$ is the inner radius of the shell, 
$r_{\rm max}$ is the outer radius of the shell, 
$r_{\rm sw}$ is the radius of the superwind 
between $r_{\rm min}$ and the AGB wind
which defines the ``thickness'' of the inner, 
axisymmetric region of the shell.
The model parameters 
$r_{\rm min}$, $r_{\rm sw}$ and $r_{\rm max}$, 
when available,
are taken from the literature
where these parameters are constrained 
by the optical and IR morphology
(see Table~\ref{tab:modpara}). 
The expansion velocity $v_{\rm exp}$
is also taken from the literature,
which was mostly determined from 
the CO line emission. 

This function defines a three-layer 
density distribution.
The outermost region has
a spherically-symmetric, power-law density
distribution $\rho(r,\Theta) \propto r^{-B}$
which results from the early AGB mass loss 
occurring in an almost perfect spherical symmetry.
The density distribution of the inner-most region
is axisymmetric and has an equatorial enhancement.
The equatorial enhancement is considered to be caused 
by the axisymmetric superwind 
at the end of the AGB phase.
The degree of the equatorial enhancement 
is controlled by $A$
(with $A=0$ corresponding to spherical symmetry,
i.e., no enhancement).
The equatorial enhancement can be made 
disk-like or toroidal-like by 
the ``flatness'' parameter $F$,
with small $F$ values for toroidal density distributions
and large $F$ values for disk-like structures.
The transitional mid-region is controlled by $C$,
the shell elongation parameter. 
It turns on and off the latitudinal dependence 
of the radial fall-off parameter, $B$.
%
The $D$ and $E$ parameters describe 
the ``abruptness'' of the geometrical transition: 
the larger $D$ and $E$ are,
the more abrupt is the dissipation of 
the latitudinal variation in the 
density distribution (see Ueta \& Meixner 2003).

For the dust composition,
we only consider amorphous carbon dust. 
The 21$\mum$ and 30$\mum$ carriers are
not included in our modeling for three reasons:
(1) their carriers are unidentified; 
(2) their excitation mechanisms are unknown,
    i.e., are they nano-sized and undergo 
     transient heating by single stellar photons
     [Draine \& 2001, Liu et al.\ 2002]
     like FeO [Li et al.\ 2013]
     and TiC nanoparticles [Li 2003]
     or sub-$\mu$m-sized and attain 
     an equilibrium temperature 
     like SiC [Jiang et al.\ 2005]
     and hydrogenated amorphous carbon 
     [HAC; Zhang et al.\ 2009a])? and
(3) we are mainly interested to 
derive the dust mass loss rates.
The carriers of the 21$\mum$ and 30$\mum$ features
are not important as far as the mass loss rates are 
concerned (see Hony et al.\ 2002, 2003, 
Hony \& Bouwman 2004).
The absorption and scattering cross sections 
are computed with Mie theory,
using the dielectric functions of amorphous carbon
of Rouleau \& Martin (1991).

For the dust size distribution, 
we take a MRN-type power-law 
distribution function of $dn/da \propto a^{-\beta}$ 
for $a_{\rm min} < a < a_{\rm max}$,
where $a$ is the spherical radius of the dust
(we assume the dust to be spherical)
with a lower and upper cutoff of
$a_{\rm min}=50\Angstrom$ and
$a_{\rm max}=0.25\mum$, respectively
(see Mathis et al.\ 1977).
For a given dust size distribution and 
a given dust shell structure, the optical depth
$\tau_\lambda$ at wavelength $\lambda$ is directly
related to the dust quantity.

Admittedly, the 2-DUST model involves 
many parameters 
(see Ueta \& Meixner 2003).
General considerations help explore parameter space.
The equatorial enhancement parameter $A$
which sets the equator-to-pole density ratio 
($\rho_{\rm eq}/\rho_{\rm pol} = 1 + A$)
is mainly constrained by the mid-IR emission
from warm dust (i.e., the {\it IRAS} 25$\mum$ photometry
of the 21$\mum$ sources). 
Generally speaking, models with high $A$ values
produce more mid-IR emission.
The radial-falloff parameter $B$ is strongly 
tied to the mass-loss dynamics: 
$B=2$ for a steady mass loss,
$B>2$ for a steadily increasing mass loss, and
$B<2$ for a diminishing mass loss
(see Ueta \& Meixner 2003).
The $B$ parameter 
is mainly constrained by the cold dust emission 
at $\lambda\gtsim 100\mum$, 
particularly by the 1.2\,mm emission.
For models with a larger $B$ value,
more dust will be concentrated radially 
closer to the central star 
and a larger amount of warmer dust 
will be present in the shell
and will therefore emit too much
at the mid-IR and too little in 
the far-IR at $\lambda\gtsim 100\mum$.
%
In view that for 12 over 18 of our sources 
there is no 1.2\,mm photometry available 
and the longest wavelength data available 
so far is the {\it IRAS} 100$\mum$ photometry.
It is therefore possible that our SED modeling 
may favor large $B$ values. 
To avoid this, we intend to choose 
the smallest $B$ value which still fits
the observed SED. 
The shell elongation parameter $C$
only slightly affects the far-IR emission:
models with a larger $C$ produces slightly
more emission at $\lambda>100\mum$.
Model SEDs with different $D$, $E$, 
and $F$ parameters also show very little 
difference in the IR.
The model parameters 
(particularly $C$, $D$, $E$ and $F$)
would be better constrained by 
the optical and IR morphology 
of the 21$\mum$ sources 
(see Ueta \& Meixner 2003).

The 21 and 30$\mum$ features could appreciably contribute
to the mid-IR broadband photometry. 
Particularly, an appreciable fraction of 
the 21.34$\mum$ {\it MSX} $E$-band flux 
of each source could arise from 
the 21$\mum$ feature. This is probably also true
for the {\it AKARI} 18$\mum$ band, 
the {\it WISE} 22$\mum$ band,
and the {\it IRAS} 25$\mum$ band. 
The broad 30$\mum$ feature could account for 
a smaller (but not necessarily negligible) fraction 
of these broadband detections.
Since the carriers of the 21 and 30$\mum$ features are not
included in our model, the contributions of these features 
to the broadband photometry need to be subtracted. 
To this end, we approximate the 21 and 30$\mum$ features
of each source by four Drude profiles respectively 
peaking at 20.1, 26, 30, and 33$\mum$ 
and integrate the sum of these Drude profiles 
with the filter functions of
the {\it AKARI} 18$\mum$ band, 
the 21.34$\mum$ {\it MSX} $E$ band,
the {\it WISE} 22$\mum$ band,
and the {\it IRAS} 25$\mum$ band. 
In Table~\ref{tab:Feat2Photo} we tabulate
the contributions of the 21 and 30$\mum$ features
of each source to the photometric fluxes of these bands.
In the following, unless otherwise stated,
the mid-IR photometric fluxes refer to
the feature-subtracted broadband photometry.
It is these feature-subtracted photometric fluxes
that we will model.

The inclination ($\theta$) of a nebula is best 
constrained by the observed nebula morphology. 
For some of our sources, the inclination angles 
have already been derived in the literature from 
their resolved optical and/or near-IR images. 
For these sources, we adopt the inclination angles 
reported in the literature
(see Table~\ref{tab:modpara}).
For the remaining sources, we consider 10 different
inclination angles increasing from $\theta=0^{\rm o}$  
(i.e., pole-on) to $\theta=90^{\rm o}$ (i.e., edge-on)
at a step of $\Delta\theta=10^{\rm o}$.
We take the inclination angle 
which provides the best fit to 
the observed SED.

\begin{figure*}
\centerline{
\includegraphics[width=15.0cm]{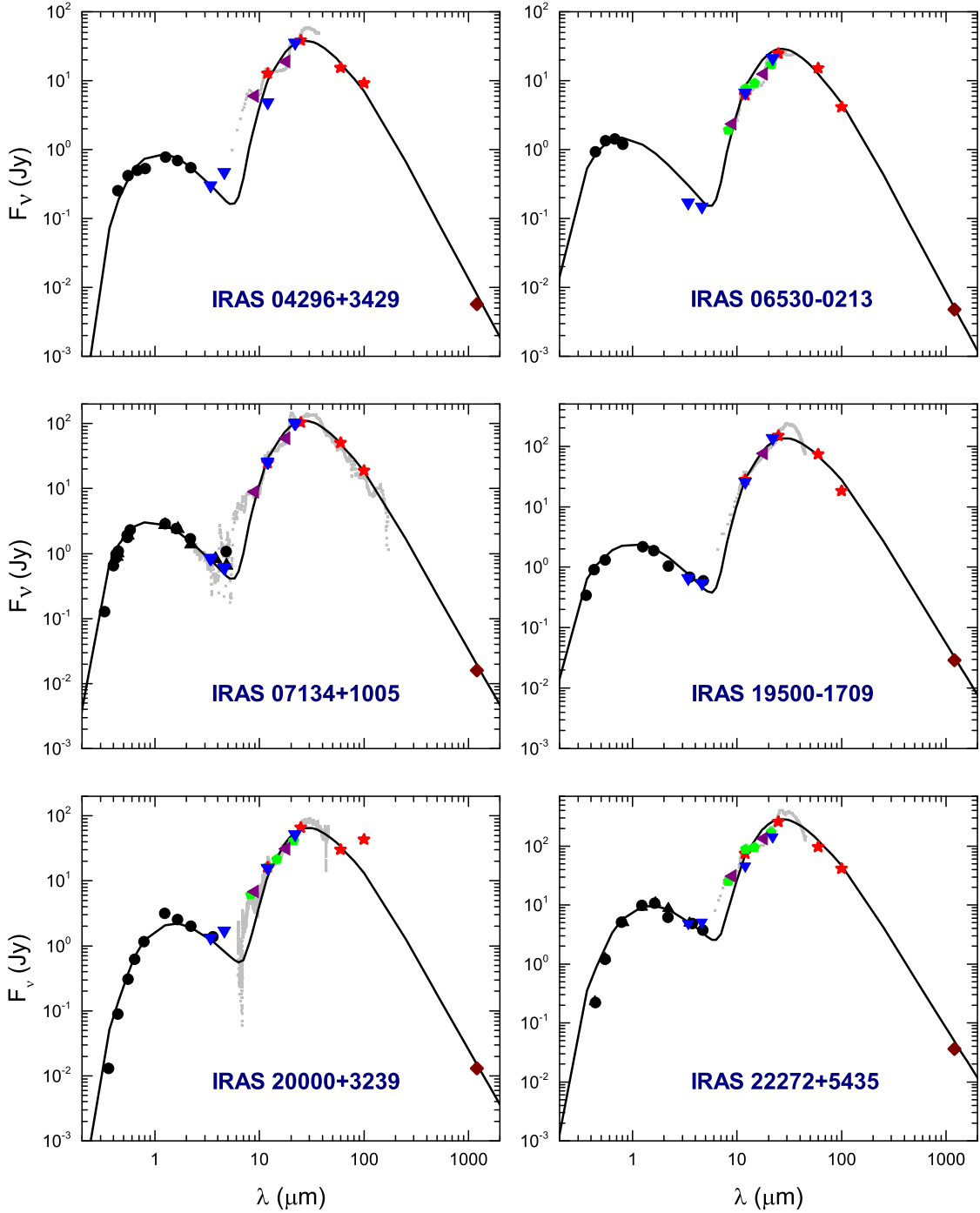}
}
\caption{\footnotesize
         \label{fig:irem1}
         Comparison of the model SEDs 
         with the broadband photometry 
         of {\it IRAS} (red stars),
         {\it AKARI} (purple triangles),
         {\it WISE} (blue triangles), 
         {\it MSX} (green pentagons), and
         the 1.2\,mm {\it IRAM}/MAMBO data (wine diamonds),
         as well as the optical/near-IR stellar 
         photospheric emission (filled circles) 
         and the {\it ISO}/SWS or {\it Spitzer}/IRS 
         mid-IR spectra (gray lines)
         for IRAS\,04296+3426, 
	 IRAS\,06530-0213, IRAS\,07134+1005, IRAS\,19500-1709, 
	 IRAS\,20000+3239, and IRAS\,.22272+5435. 
         The model SEDs were computed
         using the 2-DUST code 
         of Ueta \& Meixner (2003).
         We note that the contributions
         of the 21 and 30$\mum$ features to 
         the photometric fluxes of 
         the {\it AKARI} 18$\mum$ band, 
         the {\it MSX} $E$ band at 21.34$\mum$,
         the {\it WISE} 22$\mum$ band,
         and the {\it IRAS} 25$\mum$ band
         have been subtracted (see \S\ref{sec:massloss}). 
         }
\end{figure*}

Figure~\ref{fig:irem1} shows 
the model fits to the observed dust 
and stellar photometry
of the six Galactic sources 
for which the 1.2\,mm flux has been 
measured by IRAM/MAMBO. 
The overall fits are reasonably satisfactory.
We note that we do not intend 
to fit the ISO/{\it Spitzer} spectroscopy 
as we are mostly interested to derive 
the dust mass loss rates.  
The {\it Spitzer}/IRS or {\it ISO}/SWS spectra
overlaid in the SED fit mostly serve as a guide. 
For IRAS\,20000+3239, the model could not fit
the IRAS 100$\mum$ flux. In view of the overall SED,
it is likely that the IRAS 100$\mum$ flux could have 
been overestimated. 

\begin{figure*}
\centerline{
\includegraphics[width=15.0cm]{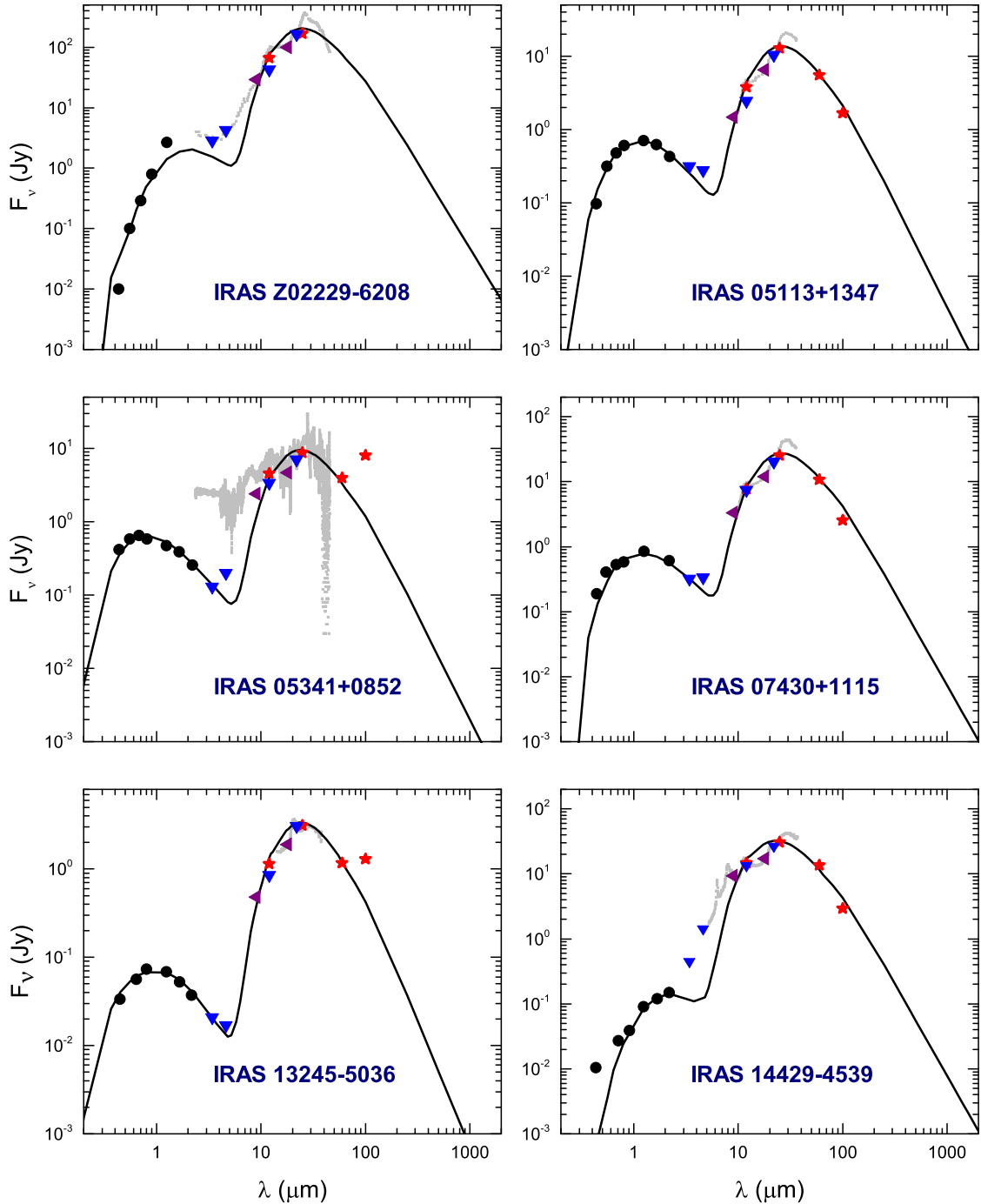}
}
\caption{\footnotesize
         \label{fig:irem2}
         Comparison of the model SEDs 
         with the broadband photometry 
         of {\it IRAS} (red stars),
         {\it AKARI} (purple triangles),
         {\it WISE} (blue triangles), 
         {\it MSX} (green pentagons), 
         as well as the optical/near-IR stellar 
         photospheric emission (filled circles) 
         and the {\it ISO}/SWS or {\it Spitzer}/IRS 
         mid-IR spectra (gray lines)
	 for IRAS\,Z02229-6208, IRAS\,05113+1347, 
	 IRAS\,05341+0852, IRAS\,07430+1115, 
	 IRAS\,13245-5036, and IRAS\,14429-4539.
         The contributions
         of the 21 and 30$\mum$ features to 
         the photometric fluxes of 
         the {\it AKARI} 18$\mum$ band, 
         the {\it MSX} $E$ band at 21.34$\mum$,
         the {\it WISE} 22$\mum$ band,
         and the {\it IRAS} 25$\mum$ band
         have been subtracted.
         }
\end{figure*}

\begin{figure*}
\centerline{
\includegraphics[width=15.0cm]{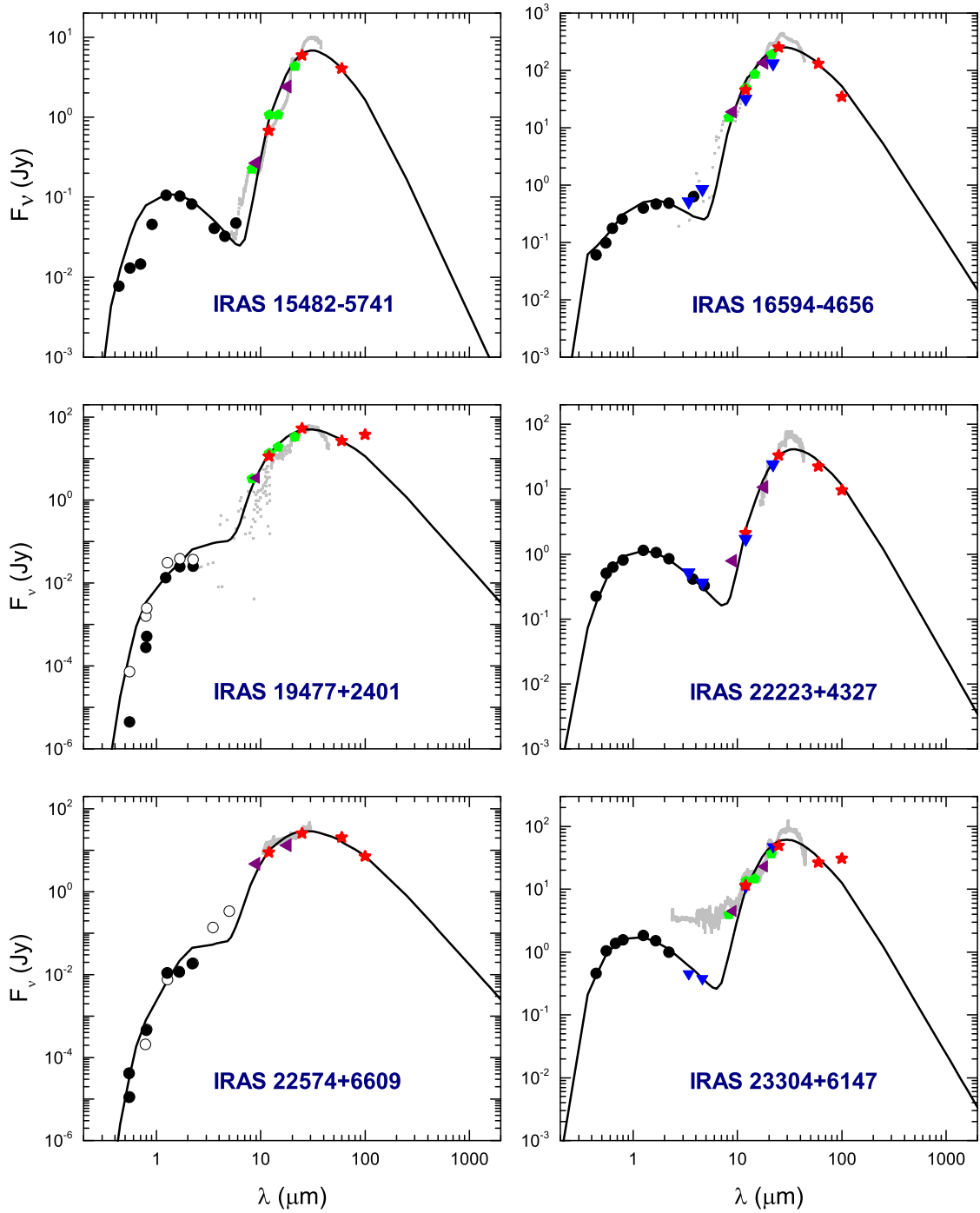}
}
\caption{\footnotesize
         \label{fig:irem3}
         Same as Figure~\ref{fig:irem2} 
         but for IRAS\,15482-5741, 
	 IRAS\,16594-4656, IRAS\,19477+2401, 
	 IRAS\,22223+4327, IRAS\,22574+6609, IRAS\, 
	 and IRAS\,23304+6147.
         }
\end{figure*}

Figures~\ref{fig:irem2} and \ref{fig:irem3}
show the model fits to the observed SEDs 
of the 12 Galactic sources 
for which there are no reported 
1.2\,mm IRAM/MAMBO measurements. 
Similarly, the model fits the observational data
fairly well, although it fails at reproducing 
the 21$\mum$ and 30$\mum$ emission features
as it is not our intention to fit these features. 
Similar to IRAS\,20000+3239,
for IRAS\,05113+1347,
IRAS\,05341+0852, IRAS\,13245-5036, 
IRAS\,19477+2401, and IRAS\,23304+6147, 
the IRAS 100$\mum$ flux could 
have been overestimated.
In order to examine 
whether the lack of any photometric data 
at $\lambda>100\mum$ would ``miss'' the dust that 
is cold and located in the outer region of the shell 
(leading to an underestimate of the dust mass), 
we also fit the SEDs of the six sources 
shown in Figure~\ref{fig:irem1} 
for which the 1.2\,mm photometric data are available,
but neglecting the 1.2\,mm photometry. 
We find that the dust mass derived from 
ignoring the 1.2\,mm data
only differs by $<$\,30\%
from that in which the 1.2\,mm data are included.
This is because, except for IRAS\,20000+3239,
the 60, 100$\mum$ and 1.2\,mm data do not 
deviate much from a single power-law 
and thus no cold component is missing.

In Table~\ref{tab:modpara} we list the model
parameters for the dust density function
($A$,  $B$,  $C$,  $D$,  $E$,  and $F$),
the dust size distribution power index $\beta$,
the expansion velocity $v_{\rm exp}$, 
and the inner ($r_{\rm min}$) 
and outer ($r_{\rm max}$) 
boundaries of the PPN dust shell.
It is seen from Table~\ref{tab:modpara} that
the best-fit models for all the Galactic 
21$\mum$ sources favor $B>2$, implying
a non-steady mass-loss process. 
As illustrated in eq.\,\ref{eq:dndr}, 
the $B$ factor also couples with 
the density distribution of the SW component.
Therefore, our best-fit models suggest
that both the AGB mass-loss and the SW mass-loss
are not constant (i.e., $B\neq2$). 
While in the literature a steady mass-loss
is often assumed for modeling the millimeter 
data of CO and even for the dust IR emission 
SED modeling, we note that a constant mass-loss 
rate is often an idealistic and simplified assumption. 
The production of the mass loss is a complicated 
process involving both the pulsation of the surface 
layers of AGB stars and the stellar radiation pressure 
on dust (see Habing et al.\ 1994). Near the end
of the AGB phase, one expects an accelerating 
increase in mass loss (see Renzini 1981, 
Gail \& Sedlmayr 2014). 
Therefore, one would not expect 
a constant mass-loss rate along the AGB 
evolution. Nevertheless, if there is only limited
observational information about the CO emission
or the dust IR emission, it is not unreasonable 
to adopt $B=2$.

With the assumption of constant velocity mass loss,
we calculate the time scale ($\tau$) for each source 
by dividing the size of the extended dust shell 
with its expansion velocity $v_{\rm exp}$. 
For the AGB phase, we obtain a time scale
of $\tauAGB = \left(\rmax-\rsw\right)/\vexp$.
For the superwind phase, the time scale is
$\tauSW = \left(\rsw-\rmin\right)/\vexp$
(see Ueta \& Meixner 2003).
Our model calculations indicate that the short duration 
of the intense mass loss from most of the post-AGB stars 
in the AGB phase to be $\sim$\,$10^3$ years 
and the superwind phase to be $\sim$\,$10^2$ years.
The dust mass for most of the sources 
in the AGB and superwind phases is in 
the range of $\sim$\,$10^{-3}$\,$M_{\odot}$.
The average mass loss rate 
for each phase ($\dMAGB$, $\dMSW$)
is calculated from the dust mass 
($\MdustAGB$, $\MdustSW$)
and the duration of mass loss 
($\tauAGB$, $\tauSW$)
assuming a gas-to-dust 
mass ratio of $\sim$\,280 
appropriate for C-rich AGB stars
(Justtanont et al.\ 1996), 
i.e., 
$\dMAGB = 280\,\MdustAGB/\tauAGB$
and 
$\dMSW = 280\,\MdustSW/\tauSW$.
In Table~\ref{tab:massloss} we tabulate 
the dust masses and mass loss rates of 
each source in the AGB and superwind phases.
Due to the inter-dependencies of the model parameters,
we estimate the uncertainties from the range of parameters 
used in our model calculations. 
We note that the actual dust masses and mass-loss rates
are expected to be somewhat larger since the carriers of
the 21 and 30$\mum$ features are not included in our SED
fitting. The carriers of these features, particularly that
of the 30$\mum$ feature, are expected to account for
an appreciable fraction of the total dust mass 
since they account for a substantial fraction 
of the total IR luminosity. 
However, without knowing the mineralogical compositions 
and optical properties of their carriers, 
we are not able to determine their masses.

\begin{figure*}
\begin{center}
\includegraphics[width=9.0cm]{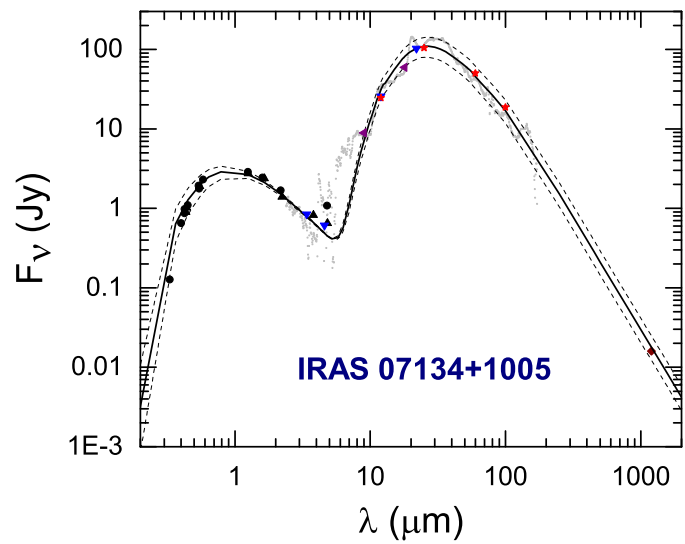}
\caption{\footnotesize
         \label{fig:error}
         Determination of the uncertainty for 
         $\tau_{9.8}$ for IRAS\,07134+1005.
         The data points plot 
         the broadband photometry 
         of {\it IRAS} (red stars),
         {\it AKARI} (purple triangles),
         {\it WISE} (blue triangles), 
         {\it MSX} (green pentagons), 
         as well as the optical/near-IR stellar 
         photospheric emission (filled circles). 
         The solid line plots the   
	 the best-fit model 
         with $\tau_{9.8}\approx0.022$,
         while the dashed lines correspond to 
         the model SEDs obtained with $\tau_{9.8}$ 
         set to the extremes of its allowable range:
         $\tau_{9.8}\approx0.0146$ 
         for the lower curve 
         and $\tau_{9.8}\approx0.0294$ 
         for upper curve. 
         }
\end{center}
\end{figure*}

To gauge the range over which a parameter could vary 
while the overall model fit
to the observed SED remains acceptable, 
following the approach of Sargent et al.\ (2010),
we allow the dust IR fluxes to 
deviate by one to three times the uncertainties, 
as estimated by eye, while keeping all other 
parameters at their best-fit values.  
Figure~\ref{fig:error} demonstrates 
this estimation of uncertainty by eye 
for the optical depth parameter 
$\tau_{9.8}$ at wavelength $\lambda=9.8\mum$ 
for IRAS\,07134+1005, 
given as dashed curves calculated from models 
with $\tau_{9.8}$ set to the extremes 
of its allowable range.
The uncertainties on the other parameters 
are determined in a similar way for all 18 sources.
%
%


Finally, we admit that, to accurately explore 
parameter space, the SED modeling itself is not
sufficient. A simultaneous modeling of the SEDs
and the resolved scattered-light images 
in the optical/near-IR as well as other data
(e.g., spectropolarimetry) would allow many parameters
to be more accurately explored 
(e.g., see Oppenheimer et al.\ 2005). 
Nevertheless, the parameters derived here 
are generally consistent with that from 
more sophisticated models. For example, 
for the common source IRAS~04296+3429, 
our mass loss rate and mass loss duration
agree with that of Oppenheimer et al.\ (2005) 
within the uncertainty range.

\section{Results and Discussion
         \label{sec:correlation}}         
We now explore the relation of the 21$\mum$ feature
with the stellar mass loss rates.
We take the flux 
emitted from the 21$\mum$ feature $\Ftwenty$ 
of ten 21$\mum$ sources 
determined by Mishra, Li \& Jiang (2015)
and that of eight sources derived 
in \S\ref{sec:sample}
(see Table~\ref{tab:drudefit} 
and Figure~\ref{tab:drudefit}).
To cancel the distance effect,
we multiply the mass loss rates by 
$\left(d/{\rm 1\,kpc}\right)^{-2}$,
where $d$ is the distance of the source to Earth.

\begin{figure*}
\begin{center}
\includegraphics[width=9.0cm]{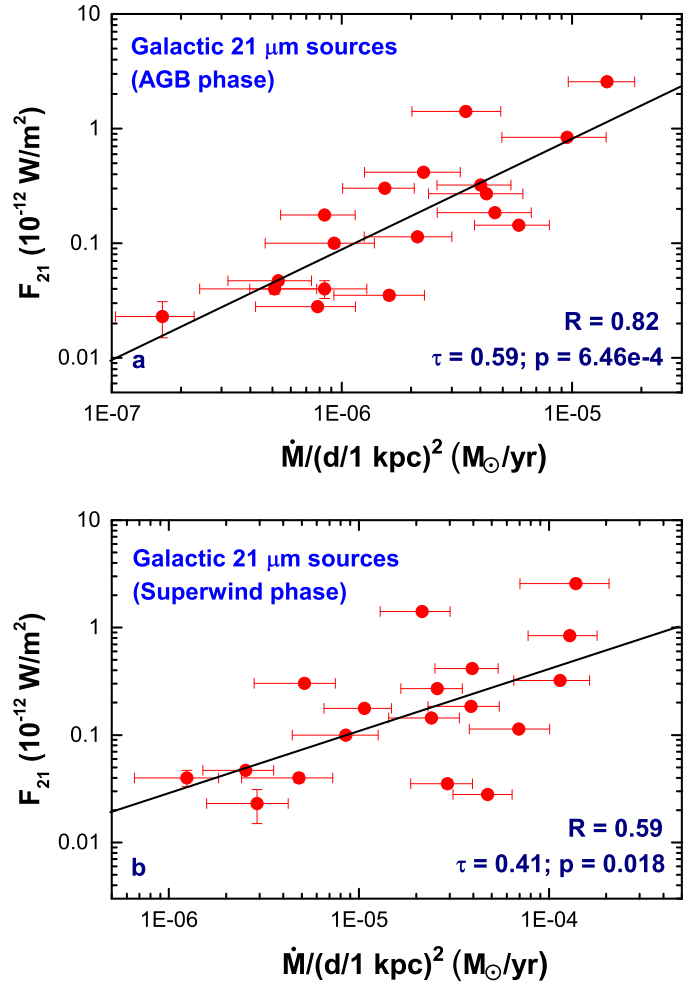}
\caption{\footnotesize
         \label{fig:21um.mass.loss}
         Correlation of the 21$\mum$ feature 
         with the stellar mass loss rates 
         in the AGB phase (a)
         and in the superwind phase (b).
         Note that the error-bar sizes of 
         $\Ftwenty$ and $\Fthirty$ are smaller 
         than that of the red circles.
         }
\end{center}
\end{figure*}

In Figure~\ref{fig:21um.mass.loss}
we plot the flux emitted from 
the 21$\mum$ feature ($\Ftwenty$)
against the stellar mass loss rates $\dot{M}$
of the AGB phase ($\dMAGB$) 
and the PPN superwind phase ($\dMSW$),
multiplied by $\left(d/{\rm 1\,kpc}\right)^{-2}$.
A linear parametric test leads to
a Pearson correlation coefficient 
of $R\approx0.82$ for all the 18 Galactic sources. 
The correlation is significant at 
a $\simali$2.1$\sigma$ level. 
We also perform a Kendall non-parametric test 
to measure the strength of dependence between 
$\Ftwenty$ and $\dMAGB$ or $\dMSW$.
We derive a Kendall correlation coefficient 
of $\tau\approx0.59$ and a corresponding probability
$p\approx6.46\times10^{-4}$ of a chance correlation 
at a 3$\sigma$ significant level.
Therefore, the 21$\mum$ feature shows 
a tendency of correlating with $\dMAGB$.
In contrast, with a Pearson correlation coefficient 
of $R\approx0.59$ and a Kendall $\tau\approx0.41$ 
and $p\approx0.018$,
the 21$\mum$ feature shows 
a much weaker correlation with $\dMSW$, if at all. 
This suggests that the carriers 
of the 21$\mum$ feature are
probably formed in the AGB phase.
%
%

%
The nondetection of the 21$\mum$ feature
in AGB stars suggests that
the excitation of its carriers
may require ultraviolet (UV) photons 
which are available in PPNe but not in AGB stars. 
More likely, the 21$\mum$ feature carriers
could be embedded in or attached to 
some sort of bulk carbon dust
(e.g., sub-$\mu$m-sized HAC) 
as nano-sized ``islands'' or ``side-groups units''. 
Upon leaving the AGB phase,
the UV photons of PPNe
break them away from the bulk dust 
and excite them to emit at 21$\mum$.
Ultimately, they are destroyed by the much harder
UV photons in PNe. This explains the exclusive
detection of the 21$\mum$ feature in PPNe.

%

The 21$\mum$ sources also emit strongly 
at the 30$\mum$ feature (Volk 2015).
While the 21$\mum$ feature is only detected in PPNe, 
the 30$\mum$ feature is more commonly seen in 
carbon-rich objects at various evolutionary stages,
spanning the AGB, PPN and PN phases
(see Jiang et al.\ 2010, Zhang \& Jiang 2008).
First detected in several carbon stars 
and in two PNe (Forrest et al.\ 1981), 
the 30$\mum$ feature 
also remains unidentified.
Magnesium sulfide (MgS) dust,
the most popular candidate carrier 
(Goebel \& Moseley 1985, Nuth et al.\ 1985,
Jiang et al.\ 1999, Szczerba et al.\ 1999, 
Hony et al.\ 2002, 2003,
Lombaert et al.\ 2012),
has recently been ruled out
as a valid carrier 
since it would require too much S 
to account for the observed fluxes
of the 30$\mum$ feature 
(see Zhang et al.\ 2009b; 
also see Messenger et al.\ 2013,
Otsuka et al.\ 2014).

\begin{figure*}
\begin{center}
\includegraphics[width=9.0cm]{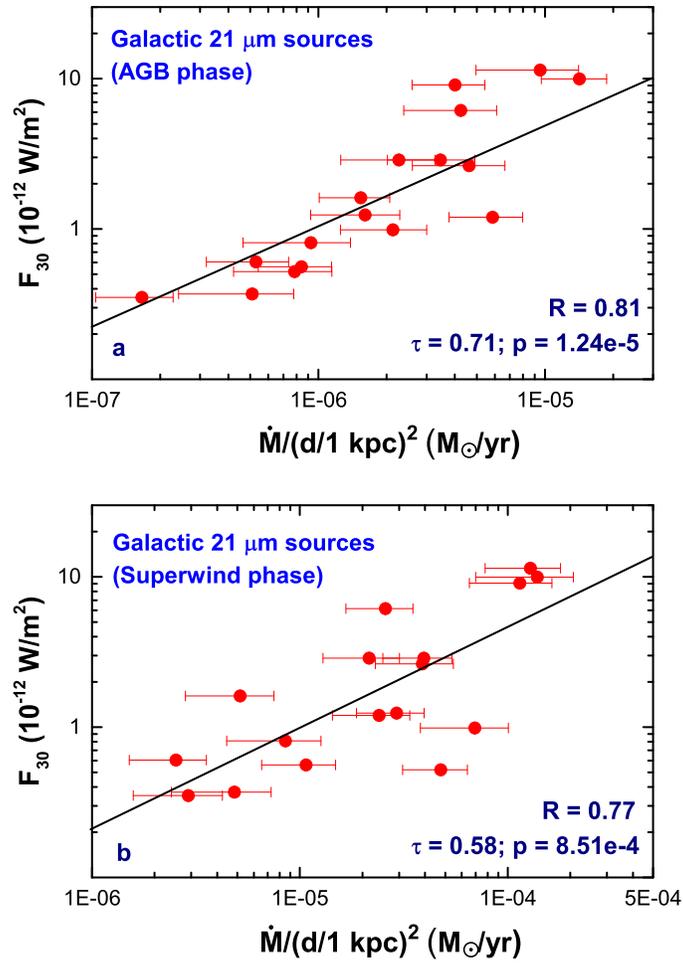}
\caption{\footnotesize
         \label{fig:30um.mass.loss}
         Correlation of the 30$\mum$ feature
         with the stellar mass loss rates 
         in the AGB (a) and superwind (b) phases. 
         }
\end{center}
\end{figure*}

We have also examined the correlation
of the 30$\mum$ feature with the stellar mass
loss rates $\dMAGB$ and $\dMSW$. 
As shown in Figure~\ref{fig:30um.mass.loss},
with a Pearson correlation coefficient 
of $R\approx0.81$
and a Kendall $\tau\approx0.71$ 
and $p\approx1.24\times10^{-5}$,
the 30$\mum$ feature shows a moderate correlation
with $\dMAGB$. This, consistent with its detection
in AGB stars, implies that its carriers condense in
the AGB phase and its excitation does not require 
UV photons. Figure~\ref{fig:30um.mass.loss} also shows
that, with $R\approx0.77$, $\tau\approx0.58$ 
and $p\approx8.51\times10^{-4}$, 
the 30$\mum$ feature and the superwind
is somewhat correlated, suggesting that the 30$\mum$
feature carriers could also condense in the superwind.
%

%

Based on the principal component analysis (PCA) method,
we have also applied a fully non-parametric multivariate 
analysis to the four variables $\Ftwenty$, $\Fthirty$, 
$\dMAGB$ and $\dMSW$ to determine their possible correlations
and statistical significance. 
As shown in Table~\ref{tab:multivariate},
the results derived from the PCA multivariate analysis
technique are in close agreement with that of
the Pearson linear parametric analysis and 
that of the Kendall non-parametric analysis:
while the 21$\mum$ feature appears to correlate with 
$\dMAGB$ but not with $\dMSW$,
the 30$\mum$ feature seems to correlate with
both $\dMAGB$ and $\dMSW$.
%
%

\begin{figure*}
\begin{center}
\includegraphics[width=9.0cm]{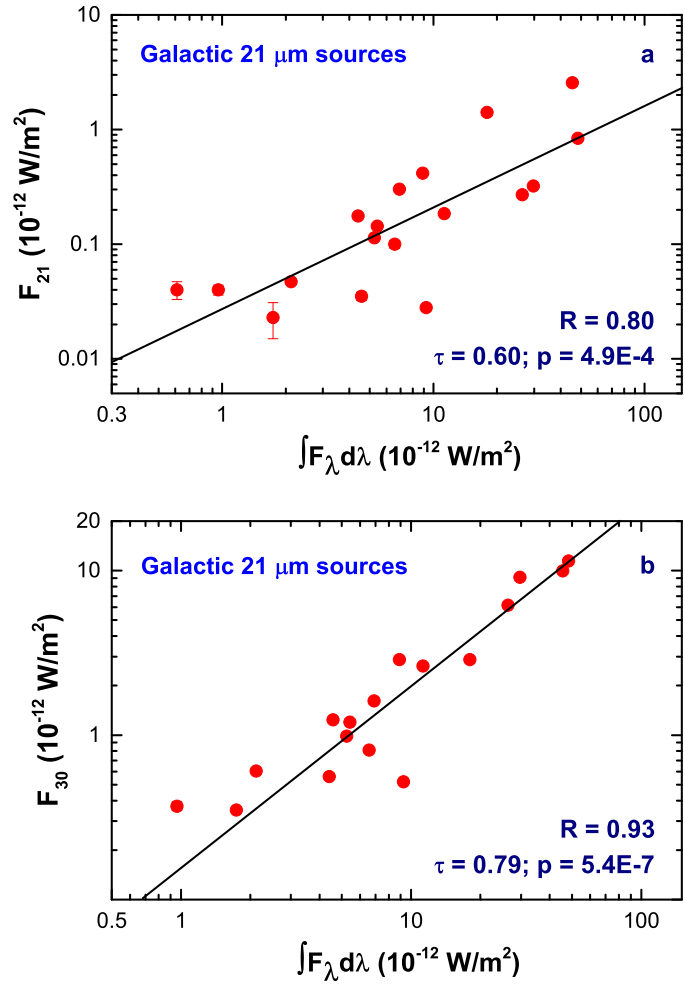}
\caption{\footnotesize
         \label{fig:21um.30um.IRtot}
         Correlation of the 21$\mum$ feature (a)
         and the 30$\mum$ feature (b)
         with the total IR emission
         obtained by integrating 
         the observed dust IR SED
         over the entire wavelength range.
         } 
         
\end{center}
\end{figure*}

In Figure~\ref{fig:21um.30um.IRtot}
we show the correlations of $\Ftwenty$ and $\Fthirty$
with the total IR emission
$\Fir \equiv \int F_\lambda\,d\lambda$
obtained by integrating the observed dust IR SED
over the entire wavelength range.
While it is apparent that both $\Ftwenty$ and $\Fthirty$
correlate with $\Fir$, the latter shows a closer
correlation with $\Fir$ than the former.  
This supports the idea of the condensation of
the 21$\mum$ feature carrier 
mainly occurring in the AGB phase 
while the 30$\mum$ feature carrier could condense
both in the AGB phase and in the superwind phase.
If one assumes that all kinds of dust species are
proportionally condensed, 
one would expect $\Ftwenty$ to be proportional 
to the IR power $\Fir^{\scriptscriptstyle\rm AGB}$ 
emitted by the bulk dust generated in the AGB phase.
In contrast, $\Fthirty$ is expected to 
be proportional to the total IR power
$\Fir = \Fir^{\scriptscriptstyle\rm AGB} 
+ \Fir^{\scriptscriptstyle\rm SW}$
emitted both by the bulk dust 
generated in the AGB phase
($\Fir^{\scriptscriptstyle\rm AGB}$)
and by the bulk dust 
generated in the superwind
($\Fir^{\scriptscriptstyle\rm SW}$).
%
%
It is worth noting that while the mass loss rates
$\dMAGB$ and $\dMSW$ derived in \S\ref{sec:massloss}
have not included the (unknown) carriers 
of the 21 and 30$\mum$ features,
the total IR emission $\Fir$ is obtained by integrating
over the entire SED and therefore does include 
the contributions from both features.
The fact that the correlations of $\Fir$ 
with $\Ftwenty$ and $\Fthirty$ are consistent
with the correlations of $\dMAGB$ 
(and $\dMSW$ for the 30$\mum$ feature as well)
with $\Ftwenty$ and $\Fthirty$
implies that the exclusion of 
the 21 and 30$\mum$ feature carriers
in deriving the mass loss rates does not
affect our correlation studies.
%

The 21$\mum$ sources also emit at
the UIR bands (Hrivnak et al.\ 2008).
%
However, the spectral profiles of the UIR bands
of the Galactic 21$\mum$ sources are ``unusual''
in the sense that they appear substantially 
different from that of the typical interstellar 
UIR bands (see Peeters et al.\ 2002). 
Most notably, while at $\simali$8$\mum$ 
the interstellar UIR bands have two well-separated 
features at 7.7 and 8.6$\mum$, the Galactic 21$\mum$ 
sources have a broad 8$\mum$ feature. 
%
%
The UIR features are widely seen in PPNe, PNe 
and the interstellar medium (ISM).
However, they are rarely seen in AGB stars.
The few C stars that display the UIR features all have 
a hot companion that emits UV photons 
(Speck \& Barlow 1997, Boersma et al.\ 2006).
In the context of PAHs as the carriers of 
the UIR features, one may ascribe the nondetection
of the UIR features in C stars to that,
due to lack of UV photons in cool C-rich AGB stars, 
PAHs, even present in the circumstellar envelopes
around C-rich AGB stars, may not be sufficiently 
excited to emit in the near- and mid-IR.
However, Li \& Draine (2002) have 
demonstrated that the excitation of PAHs 
does not require UV photons, and
the visible/near-IR photons available 
in C stars are capable of exciting PAHs
to emit at the ``UIR'' bands.
The visible/near-IR absorption spectra 
measured by Mattioda et al.\ (2005) for 
PAH ions further support the finding of
Li \& Draine (2002) that PAHs can be excited 
by the soft stellar photons from C stars. 

In principle, it would be of great value 
to examine the correlation
of the UIR features with the stellar mass
loss rates $\dMAGB$ and $\dMSW$.
{\it If} the UIR features are shown to correlate 
with $\dMAGB$, one could speculate that their carriers,
like that of the 21$\mum$ feature,
could form in the AGB phase as aromatic ``islands''
embedded in bulk HAC dust.
The nondetection of the UIR bands in AGB stars
could merely indicate that the carriers of 
the UIR features could not be knocked off 
the bulk dust by the photons of AGB stars 
as free-flying aromatic hydrocarbon molecules.
Unfortunately, the {\it Spitzer}/IRS 
or {\it ISO}/SWS spectra of the 21$\mum$ sources
do not always span all the UIR bands.
More specifically, for some sources 
there lack spectroscopic data at $\lambda<10\mum$ 
(e.g., see Figure~\ref{fig:drudefit})
which prevents an accurate determination 
of $\Fuir$. Therefore, we do not intend
to explore the possible correlation between
$\Fuir$ and $\dMAGB$ or $\dMSW$.
Finally, we note that 
we have not considered the nine 21$\mum$ sources 
detected in the Large Magellanic Cloud (LMC)
and the Small Magellanic Cloud (SMC)
both of which are metal-poor: 
the metallicity of the LMC is $\sim$1/4 of
that in the Galaxy (Russell \& Dopita 1992),
while the metallicity of the SMC is only 
$\simali$1/10 of that of the Galaxy 
(Kurt \& Dufour 1998).
With respect to the overall mid-IR emission
and the UIR emission, 
the relative strengths of 
the 21 and 30$\mum$ features 
of the Magellanic Cloud sources 
are appreciably weaker than 
that of the Galactic 21$\mum$ sources.
Also, the spectral appearance of the UIR features
of the Magellanic Cloud 21$\mum$ sources
is remarkably different from that of the Galaxy.
While most of the Galactic 21$\mum$ sources have 
``unusual'' UIR spectral profiles, 
the Magellanic Cloud 21$\mum$ sources 
show more ``normal'' looking UIR features. 
Therefore, we prefer not to investigate the relation
between $\Fuir$ and $\dMAGB$ or $\dMSW$ of the Galactic
and Magellanic Cloud 21$\mum$ sources as one class.
The UIR carriers of the Galactic 21$\mum$ sources 
could be richer in aliphatics than that of the ISM
and the Magellanic Cloud 21$\mum$ sources
(see Li \& Draine 2012, Yang et al.\ 2013). 
%

%


\section{Summary\label{sec:conclusion}}
We have modeled the dust IR emission SEDs
of the 21$\mum$ sources using the 2-DUST 
radiative transfer code for 
axisymmetric dusty systems. 
We have derived their mass loss rates 
in the AGB and superwind phases. 
%
We have explored the correlation between 
the mass loss rates and the unidentified
21$\mum$ and 30$\mum$ features 
seen in carbon-rich PPNe. 
The principal results of 
this paper are the following:
\begin{enumerate}
\item The 21$\mum$ feature which is only seen 
      in carbon-rich PPNe tends to correlate
      with $\dMAGB$ but not with $\dMSW$, 
      suggesting that its carrier could condense
      in the AGB phase but its excitation and/or
      generation requires UV photons which are available
      in PPNe but not in AGB stars.
      The 21$\mum$ feature carrier could be destroyed
      by the more energetic photons available in PNe.
\item The 30$\mum$ feature which is seen in
      AGB stars, PPNe and PNe correlates with $\dMAGB$ 
      and, to a less degree, with $\dMSW$,
      suggesting that its carrier could condense
      both in the AGB phase and in the superwind
      and it can be excited by visible/near-IR photons.  
%
%
\end{enumerate}

\section*{Acknowledgements}
We thank Archana Mishra, Angela Speck, Ke Zhang 
and the anonymous referee
for their valuable comments and discussions.
We are supported in part by
NSF AST-1311804,
NNX13AE63G, 
NSFC\,11273022, NSFC\,11473023,
and the University of Missouri Research Board.



\appendix
\section{Comments on Individual Sources}
We comment on the individual sources,
focusing on the mass loss rates $\Mloss$
reported in the literature.
We will see in the following that
the mass loss rate determinations
are complicated by
(i) the unknown gas-to-dust mass ratio
$\gastodust$ if one derives $\Mloss$
from the dust emission modeling, and
(ii) the unknown CO-to-H$_2$ number density
ratio $\cotoHH$ if  one derives $\Mloss$
from the CO emission lines.
Nevertheless, the mass loss rates
derived in this work are generally
consistent with that reported in
the literature.
To facilitate comparison between 
our mass loss rates with that reported
in the literature, we correct for 
the distance and gas-to-dust mass ratio dependencies 
(i.e., $\Mloss \propto d^{2} \gastodust$).
See Table~\ref{tab:massloss} for a summary.

{\it IRAS Z02229+6208}:
This is a cool, highly reddened post-AGB star.
It has an elliptically extended nebula as revealed
by the polarization map of Ueta et al.\ (2005).
We derive a mass loss rate of
$\Mloss$\,$\approx$\,$1.94\times10^{-5}\msun\yr^{-1}$
in the AGB phase.
Hrivnak et al.\ (2000) modeled
the 2--45$\mum$ ISO spectrum of this source
using the DUSTCD radiative transfer code
of Leung (1976). Assuming a gas-to-dust
mass ratio of $\gastodust = 330$,
they derived
$\Mloss/\left\{\left(v_{\rm exp}/\km\s^{-1}\right)\times
\left(d/\kpc\right)\right\}\approx 1.3\times10^{-6}
\msun\yr^{-1}$, corresponding to a mass loss rate of
$\Mloss\approx 3.59\times10^{-5}\msun\yr^{-1}$
for $d=2.2\kpc$, $\vexp=14.8\km\s^{-1}$, 
and $\gastodust = 280$
which are adopted in this work.
Hrivnak \& Bieging (2005) estimated
$\Mloss$\,$\approx$\,$1.4\times10^{-4}\msun\yr^{-1}$
from the CO $J=4-3$ and $J=2-1$ emission lines
observed with
the {\it Heinrich Hertz Telescope} (HHT)\footnote{%
  Hrivnak \& Bieging (2005) derived
  the CO density distribution to be
  $\rhoco(r) \propto r^{-3}$.
  They adopted a number-density ratio
  of CO to H$_2$ of
  $\cotoHH=7.4\times10^{-4}$
  and assumed that all of the hydrogen
  is in the molecular form.
  In the following paragraphs,
  unless otherwise stated,
  all of the hydrogen
  in the 21$\mum$ sources is assumed to
  be in H$_2$.
  }
at $d=2.2\kpc$.

{\it IRAS 04296+3429}:
This source is in the advanced post-AGB evolution stage
with a bipolar lobe structure (Sahai 1999).
Using 2-DUST, we derive a mass loss rate of
$\Mloss$\,$\approx$\,$3.84\times10^{-5}\msun\yr^{-1}$
in the AGB phase.
Bakker et al.\ (1997) obtained the optical high
resolution absorption spectra of C$_2$ and CN.
They derived
$\Mloss$\,$\approx$\,$1.6\times10^{-6}\msun\yr^{-1}$
from C$_2$
and
$\Mloss$\,$\approx$\,$6.3\times10^{-6}\msun\yr^{-1}$
from CN.\footnote{%
  \label{ftnt:C2CN}
  They took the ratio of the number density
  of C$_2$ and CN to that of H$_2$ to be
  $\cctoHH=4\times10^{-6}$,
  $\cntoHH=3\times10^{-6}$,
  respectively.
  }
Meixner et al.\ (1997) modeled the 9.7 and 11.8$\mum$
images and the optical and IRAS photometry of this object,
using a 2-dimensional axially symmetric dust code
and amorphous carbon dust of a single size of
$a=0.01\mum$. They derived
$\Mloss$\,$\approx$\,$1.3\times10^{-5}\msun\yr^{-1}$
for $d= 4\kpc$ and $\gastodust \approx 222$
(Jura 1986). This becomes
$\Mloss\approx 3.1\times10^{-5}\msun\yr^{-1}$
for our adopted $d=5.5\kpc$
and $\gastodust = 280$.
Sahai (1999) obtained the scattered light images
of this source at 0.56 and 0.81$\mum$ using the
{\it Wide Field Planetary Camera 2} (WFPC2)
on board the {\it Hubble Space Telescope} (HST).
They derived a mass loss rate
of $\Mloss \approx 8\times10^{-6}\msun\yr^{-1}$
for $d= 4.0\kpc$ and $\gastodust \approx 200$,
which corresponds to
$\Mloss \approx$\,$1.75\times10^{-5}\msun\yr^{-1}$
if our $d=5.0\kpc$ and $\gastodust = 280$
are adopted.
Bujarrabal et al.\ (2001) derived
$\Mloss$\,$\approx$\,$4.0\times10^{-5}\msun\yr^{-1}$
based on the CO $J=1-0$ and $J=2-1$ emission lines.
Hrivnak \& Bieging (2005) estimated
$\Mloss$\,$\approx$\,$3.6\times10^{-5}\msun\yr^{-1}$
for $d=5.4\kpc$ 
from the CO $J=2-1$ emission line
observed with {\it HHT},
corresponding to
$\Mloss$\,$\approx 3.1\times10^{-5}\msun\yr^{-1}$
for our adopted $d=5.0\kpc$.  

{\it IRAS 05113+1347}:
This object is a carbon-rich G8Ia post-AGB star
(Hrivnak 1995).
We derive a mass loss rate of
$\Mloss$\,$\approx$\,$2.59\times10^{-5}\msun\yr^{-1}$
in the AGB phase.
Bakker et al.\ (1997) derived
$\Mloss$\,$\approx$\,$7.9\times10^{-7}\msun\yr^{-1}$
from C$_2$
and
$\Mloss$\,$\approx$\,$7.9\times10^{-6}\msun\yr^{-1}$
from CN (see Footnote~\ref{ftnt:C2CN}).
Hrivnak et al.\ (2009) modeled
the IR photometry up to $\lambda=100\mum$
obtained by {\it MSX} and {\it IRAS}
and the 10--36$\mum$ {\it Spitzer}/IRS spectrum
of this source using DUSTCD.
Assuming $\gastodust = 330$ and adopting $d=7.0\kpc$,
they derived
$\Mloss\approx 3.2\times10^{-4}\msun\yr^{-1}$,
corresponding to a mass loss rate of
$\Mloss\approx 2.7\times10^{-4}\msun\yr^{-1}$
for our adopted $d=7.0\kpc$  
and $\gastodust = 280$.
Reddy \& Parthasarathy (1996) derived
$\Mloss$\,$\approx$\,$4.6\times10^{-7}\msun\yr^{-1}$
for $d=5.0\kpc$
from the {\it IRAS} 60$\mum$ flux. 
This corresponds to 
$\Mloss$\,$\approx$\,$9.0\times10^{-7}\msun\yr^{-1}$
for our $d=7.0\kpc$.

{\it IRAS 05341+0852}:
This source has a very extended, 
optically thin circumstellar envelope.
Its visible image shows an elongated elliptical nebula
around the central star (Ueta et al.\ 2000).
We derive a mass loss rate of
$\Mloss$\,$\approx$\,$1.01\times10^{-5}\msun\yr^{-1}$
in the AGB phase.
Bakker et al.\ (1997) derived
$\Mloss$\,$\approx$\,$1.0\times10^{-6}\msun\yr^{-1}$
from C$_2$
and
$\Mloss$\,$\approx$\,$1.0\times10^{-5}\msun\yr^{-1}$
from CN.
Hrivnak et al.\ (2009) modeled
the optical and IR photometry
and the 10--36$\mum$ Spitzer/IRS spectrum
of this source using DUSTCD.
Assuming $\gastodust = 330$ and adopting $d=8.2\kpc$,
they derived
$\Mloss\approx 1.0\times10^{-4}
\msun\yr^{-1}$,
corresponding to a mass loss rate of
$\Mloss\approx 7.68\times10^{-5}\msun\yr^{-1}$
for our adopted $d=7.8\kpc$ 
and $\gastodust = 280$.
Reddy \& Parthasarathy (1996) derived
$\Mloss$\,$\approx$\,$2.7\times10^{-7}\msun\yr^{-1}$
from the {\it IRAS} 60$\mum$ flux
for $d=10.0\kpc$. This becomes 
$\Mloss$\,$\approx$\,$1.6\times10^{-7}\msun\yr^{-1}$
with our $d=7.8\kpc$.
%

{\it IRAS 06530-0213}:
This object is a carbon rich F15Ib post-AGB star
with a metal poor environment (Hrivnak \& Reddy 2003).
Its {\it HST} optical scattered-light image
shows an elliptical reflection nebula
(Ueta et el.\ 2000).
We derive a mass loss rate of
$\Mloss$\,$\approx$\,$1.86\times10^{-5}\msun\yr^{-1}$
in the AGB phase.
Reddy \& Parthasarathy (1996) derived
$\Mloss$\,$\approx$\,$1.8\times10^{-7}\msun\yr^{-1}$
from the {\it IRAS} 60$\mum$ flux for $d=3.0\kpc$.
This corresponds to 
$\Mloss$\,$\approx$\,$4.42\times10^{-7}\msun\yr^{-1}$ 
for our $d=4.7\kpc$.
Hrivnak et al.\ (2009) modeled
the {\it MSX} and {\it IRAS} photometry
and the 10--36$\mum$ Spitzer/IRS spectrum
of this source using DUSTCD.
Assuming $\gastodust = 330$ and adopting $d=5.4\kpc$,
they derived
$\Mloss\approx 1.2\times10^{-4}
\msun\yr^{-1}$,
corresponding to a mass loss rate of
$\Mloss\approx 7.71\times10^{-5}\msun\yr^{-1}$
for our adopted $d=4.7\kpc$ 
and $\gastodust = 280$.
Hrivnak \& Bieging (2005) estimated
$\Mloss$\,$\approx$\,$6.9\times10^{-5}\msun\yr^{-1}$
from the CO $J=2-1$ emission line
observed with {\it HHT} for $d=6.8\kpc$.
This corresponds to $\Mloss$\,$\approx$\,
$3.3\times10^{-5}\msun\yr^{-1}$ when
our $d=4.7\kpc$ is adopted.

{\it IRAS 07134+1005 (HD 56126)}:
This source is one of the best studied post-AGB star.
The circumstellar envelope around this star
has an axial symmetric structure
(Meixner et al.\ 1997; Ueta et al.\ 2000).
It is one of the sources
in which the 21$\mum$ and 30$\mum$ features
were first discovered
(Forrest et al.\ 1981, Kwok et al.\ 1989).
Our model fits all the observed SED
from the optical to the millimeter
with 2-DUST. The 2-DUST model gives
a mass loss rate of
$\Mloss$\,$\approx$\,$1.99\times10^{-5}\msun\yr^{-1}$
in the AGB phase.
Assuming $\cotoHH = 1\times10^{-3}$,
Omont et al.\ (1993) estimated
$\Mloss$\,$\approx$\,$9.7\times10^{-6}\msun\yr^{-1}$
at d=2.4 kpc from the CO $J=2-1$ line.
Hrivnak et al.\ (2000) modeled
the 2--45$\mum$ ISO spectrum of
this source using DUSTCD.
Assuming $\gastodust = 330$,
they derived
$\Mloss/\left\{\left(v_{\rm exp}/\km\s^{-1}\right)\times
\left(d/\kpc\right)\right\}\approx 7.2\times10^{-7}
\msun\yr^{-1}$,
corresponding to a mass loss rate of
$\Mloss\approx 1.47\times10^{-5}\msun\yr^{-1}$
for our adopted $d=2.4\kpc$, 
$\vexp=10.0\km\s^{-1}$, 
and $\gastodust = 280$.
Hony et al.\ (2003) modeled the optical and {\it IRAS}
photometry and the {\it ISO}/SWS and {\it ISO}/LWS spectra
of this object using the MODUST radiative transfer
code (Bouwman et al.\ 2000). They derived
$\Mloss$\,$\approx$\,$1\times10^{-4}\msun\yr^{-1}$
for $\gastodust=220$ and $d=2.4\kpc$. 
This corresponds to 
$\Mloss$\,$\approx$\,$1.2\times10^{-4}\msun\yr^{-1}$
for our $\gastodust = 280$ at $d=2.4\kpc$.
Meixner et al.\ (2004) imaged the circumstellar
envelope of this object at the CO $J=1-0$ line
using the Berkeley-Illinois-Maryland Association
(BIMA) millimeter array. They modeled the CO $J=1-0$
BIMA images and the CO $J=2-1$ (Knapp et al.\ 1998)
and $J=4-3$ (Knapp et al.\ 2000) line profiles,
using the radiative transfer code of
Justtanont et al.\ (1994).
Assuming $\cotoHH=9.2\times10^{-4}$,
they derived
$\Mloss$\,$\approx$\,$5.1\times10^{-6}\msun\yr^{-1}$
for the AGB wind.
They also modeled the optical and IRAS photometry
and the ISO spectrum of this source with 2-DUST.
They derived a dust mass loss rate of
$\Mdloss$\,$\approx$\,$9.6\times10^{-8}\msun\yr^{-1}$
for the AGB wind. If we assume a gas-to-dust ratio
of 280 (Justtanont et al.\ 1996),
the AGB wind mass loss rate would be
$\Mloss$\,$\approx$\,$2.7\times10^{-5}\msun\yr^{-1}$,
higher than that derived from CO by
a factor of $\simali$5.3.
Hrivnak \& Bieging (2005) estimated
$\Mloss$\,$\approx$\,$2.3\times10^{-5}\msun\yr^{-1}$
from the CO $J=4-3$ and $J=2-1$ emission lines
observed with {\it HHT}.
Buemi et al.\ (2007) used DUSTY to
model the SED from the optical to the millimeter
and derived
$\Mloss$\,$\approx$\,$3.48\times10^{-5}\msun\yr^{-1}$
for $\gastodust = 220$ at $d=2.4\kpc$.
This becomes 
$\Mloss\approx4.43\times10^{-5}\msun\yr^{-1}$ 
for our adopted $\gastodust=280$ 
also at $d=2.4\kpc$.
%
%

{\it IRAS 07430+1115}:
This object shows an approximately centrosymmetric
pattern as revealed by the near-IR $J$- and $K$-band
polarimetric images obtained with the 3.8\,m
{\it United Kingdom Infrared Telescope}
(UKIRT; Gledhill 2005).
We derive a mass loss rate of
$\Mloss$\,$\approx$\,$5.78\times10^{-5}\msun\yr^{-1}$
in the AGB phase.
Hrivnak et al.\ (2009) modeled
the optical and IR photometry
(of {\it MSX} and {\it IRAS})
and the 10--36$\mum$ Spitzer/IRS spectrum
of this source using DUSTCD.
With $\gastodust = 330$ and $d=6.7\kpc$,
they derived
$\Mloss\approx 1.1\times10^{-3}
\msun\yr^{-1}$,
corresponding to a mass loss rate of
$\Mloss\approx 7.48\times10^{-4}\msun\yr^{-1}$
for our adopted $d=6.0\kpc$ 
and $\gastodust = 280$.
%

%

{\it IRAS 14429-4539}:
We derive a mass loss rate of
$\Mloss$\,$\approx$\,$3.22\times10^{-5}\msun\yr^{-1}$
in the AGB phase.
Reddy \& Parthasarathi (1996) estimated
the mass loss rate to be
$\Mloss$\,$\approx$\,$3.8\times10^{-7}\msun\yr^{-1}$
for $d=6.0\kpc$
based on the simple formula of Reimers (1975):
$\Mloss \propto L_\star^{1.5}\,T_\star^{-2}\,
M_\star^{-1}$.
%

%

{\it IRAS 16594-4656}:
This object is a bipolar post-AGB star
as indicated by its SED
(Meixner et al.\ 1999, Hrivnak et al.\ 2008).
It has an optically thick circumstellar envelope.
We derive a mass loss rate of
$\Mloss$\,$\approx$\,$8.17\times10^{-5}\msun\yr^{-1}$
in the AGB phase.
Hrivnak et al.\ (2000) modeled
the 2--45$\mum$ ISO spectrum of this source
using DUSTCD.
Assuming $\gastodust = 330$
they derived
$\Mloss/\left\{\left(v_{\rm exp}/\km\s^{-1}\right)\times
\left(d/\kpc\right)\right\}\approx3.9\times10^{-6}
\msun\yr^{-1}$,
corresponding to a mass loss rate of
$\Mloss\approx 1.27\times10^{-4}\msun\yr^{-1}$
for our adopted $d=2.4\kpc$, 
$\vexp=16.0\km\s^{-1}$, 
and $\gastodust = 280$.
%


{\it IRAS 19477+2401}:
This source appears as a bipolar reflection
nebula in the {\it UKIRT} $J$-band polarimetric
image of scattered-light (Gledhill et al.\ 2001).
We derive a mass loss rate of
$\Mloss$\,$\approx$\,$2.37\times10^{-5}\msun\yr^{-1}$
in the AGB phase.
Hrivnak et al.\ (2000) modeled
the 2--45$\mum$ ISO spectrum of this source
using DUSTCD.
Assuming $\gastodust = 330$,
they derived
$\Mloss/\left\{\left(v_{\rm exp}/\km\s^{-1}\right)\times
\left(d/\kpc\right)\right\}\approx2.5\times10^{-6}
\msun\yr^{-1}$,
corresponding to a mass loss rate of
$\Mloss\approx 1.52\times10^{-4}\msun\yr^{-1}$
for our adopted $d=5.5\kpc$, 
$\vexp=13.0\km\s^{-1}$, 
and $\gastodust = 280$.

{\it IRAS 19500-1709}:
This object shows a bipolar structure
in the $J$- and $K$-band scattered-light polarimetric images
of {\it UKIRT} (Gledhill et al.\ 2001).
We derive a mass loss rate of
$\Mloss \approx 7.14\times10^{-5}\msun\yr^{-1}$
in the AGB phase.
Assuming $\cotoHH = 1\times10^{-3}$,
Likkel et al.\ (1991) derived
$\Mloss/\left(d/\kpc\right)^{2}\approx1.2\times10^{-6}
\msun\yr^{-1}$,
corresponding to a mass loss rate of
$\Mloss\approx 2.02\times10^{-5}\msun\yr^{-1}$
for our adopted $d=4.1\kpc$.
Meixner et al.\ (1997) modeled the 9.7 and 11.8$\mum$
images and the optical and IRAS photometry of this object,
using a 2-dimensional axially symmetric dust code
and amorphous carbon dust of a single size of
$a=0.01\mum$. They derived
$\Mloss$\,$\approx$\,$1.3\times10^{-5}\msun\yr^{-1}$
for $d= 2\kpc$ and $\gastodust \approx 222$
(Jura 1986). This becomes
$\Mloss\approx 6.89\times10^{-5}\msun\yr^{-1}$
for our adopted $d=4.1\kpc$
and $\gastodust = 280$.
Assuming $\cotoHH = 1\times10^{-3}$,
Omont et al.\ (1993) estimated
$\Mloss$\,$\approx$\,$8.0\times10^{-6}\msun\yr^{-1}$
at $d=2.1\kpc$ from the CO $J=2-1$ line. 
This becomes
$\Mloss\approx 3.05\times10^{-5}\msun\yr^{-1}$
for our adopted $d=4.1\kpc$.
Assuming $\gastodust = 223$, 
Clube \& Gledhill (2004) derived
$\Mloss$\,$\approx$\,$6.8\times10^{-5}\msun\yr^{-1}$
for $d= 4.0\kpc$ from fitting 
the near-, mid- and far-IR emission 
of this source. This becomes
$\Mloss\approx 8.97\times10^{-5}\msun\yr^{-1}$
for our adopted $d=4.1\kpc$
and $\gastodust = 280$.
Hrivnak \& Bieging (2005) estimated
$\Mloss$\,$\approx$\,$3.2\times10^{-5}\msun\yr^{-1}$
for $d=2.4\kpc$ 
from the CO $J=2-1$ emission line
observed with {\it HHT},
corresponding to
$\Mloss$\,$\approx 9.33\times10^{-5}\msun\yr^{-1}$
for our adopted $d=4.1\kpc$.

{\it IRAS 20000+3239}:
The near-IR imaging polarimetry of
this object clearly reveals an extended
and axisymmetric bipolar structure
(Gledhill et al.\ 2001).
Our model estimates
$\Mloss$\,$\approx$\,$2.32\times10^{-5}\msun\yr^{-1}$
in the AGB phase.
Assuming $\cotoHH = 1\times10^{-3}$,
Likkel et al.\ (1991) derived
$\Mloss/\left(d/\kpc\right)^{2}\approx1.1\times10^{-6}
\msun\yr^{-1}$,
corresponding to a mass loss rate of
$\Mloss\approx 5.52\times10^{-6}\msun\yr^{-1}$
for our adopted $d=2.24\kpc$.
Hrivnak et al.\ (2000) modeled
the 2--45$\mum$ ISO spectrum of this source
using DUSTCD.
Assuming $\gastodust = 330$
they derived
$\Mloss/\left\{\left(v_{\rm exp}/\km\s^{-1}\right)\times
\left(d/\kpc\right)\right\}\approx5.1\times10^{-7}
\msun\yr^{-1}$,
corresponding to a mass loss rate of
$\Mloss\approx 1.16\times10^{-5}\msun\yr^{-1}$
for our adopted $d=2.24\kpc$, 
$\vexp=12.0\km\s^{-1}$, 
and $\gastodust = 280$.
Buemi et al.\ (2007) used DUSTY to
model the SED from the optical to the millimeter
and derived
$\Mloss$\,$\approx$\,$6.92\times10^{-6}\msun\yr^{-1}$
for $\gastodust = 220$.
This becomes 
$\Mloss\approx3.47\times10^{-5}\msun\yr^{-1}$ 
for our adopted $\gastodust=280$ 
and $d=2.24\kpc$.
Assuming $\cotoHH = 1\times10^{-3}$,
Omont et al.\ (1993) estimated
$\Mloss$\,$\approx$\,$1.2\times10^{-5}\msun\yr^{-1}$
at $d=5.5\kpc$ from the CO $J=2-1$ line. 
This becomes
$\Mloss\approx 1.99\times10^{-6}\msun\yr^{-1}$
for our adopted $d=2.24\kpc$.

{\it IRAS 22223+4327}:
The near-IR imaging polarimetry of
this object reveals an extended scattering
envelope that is optically thick and
illuminated by the central star
(Gledhill et al.\ 2001).
We derive a mass loss rate of
$\Mloss$\,$\approx$\,$2.18\times10^{-5}\msun\yr^{-1}$
in the AGB phase.
Assuming $\cotoHH = 1\times10^{-3}$,
Likkel et al.\ (1991) derived
$\Mloss/\left(d/\kpc\right)^{2}\approx6.0\times10^{-7}
\msun\yr^{-1}$,
corresponding to a mass loss rate of
$\Mloss\approx 6.14\times10^{-6}\msun\yr^{-1}$
for our adopted $d=3.2\kpc$.
Bakker et al.\ (1997) derived
$\Mloss$\,$\approx$\,$2.5\times10^{-5}\msun\yr^{-1}$
from C$_2$
and
$\Mloss$\,$\approx$\,$1.6\times10^{-4}\msun\yr^{-1}$
from CN (see Footnote~\ref{ftnt:C2CN}).

{\it IRAS 22272+5435}:
This source is extremely carbon-rich
and fairly bright both in the IR and optical
(Ueta et al.\ 2001).
We derive
$\Mloss$\,$\approx$\,$2.59\times10^{-6}\msun\yr^{-1}$
for the AGB wind.
Bujarrabal et al.\ (2001) derived
$\Mloss$\,$\approx$\,$2.0\times10^{-5}\msun\yr^{-1}$
based on the CO $J=1-0$ and $J=2-1$ emission lines.
for $d= 1.7\kpc$. This becomes
$\Mloss\approx 1.88\times10^{-5}\msun\yr^{-1}$
for our adopted $d=1.65\kpc$.
Hrivnak \& Bieging (2005) estimated
$\Mloss$\,$\approx$\,$7.7\times10^{-5}\msun\yr^{-1}$
for $d=1.9\kpc$ 
from the CO $J=2-1$ emission line
observed with {\it HHT},
corresponding to
$\Mloss$\,$\approx 5.8\times10^{-5}\msun\yr^{-1}$
for our adopted $d=1.65\kpc$.  
Buemi et al.\ (2007) used DUSTY to
model the SED from the optical to the millimeter
and derived
$\Mloss$\,$\approx$\,$3.12\times10^{-5}\msun\yr^{-1}$
for $\gastodust = 220$.
This becomes 
$\Mloss\approx4.2\times10^{-5}\msun\yr^{-1}$ 
for our adopted $\gastodust=280$ 
and $d=1.65\kpc$.

{\it IRAS 22574+6609}:
This object is very faint in the visible.
It shows a bipolar morphology
with a dark lane dividing the nebula
into two lobes (Ueta et al.\ 2000; Su et al.\ 2001).  
We derive a mass loss rate of
$\Mloss$\,$\approx$\,$3.39\times10^{-4}\msun\yr^{-1}$
in the AGB phase.
Assuming $\cotoHH = 1\times10^{-3}$,
Likkel et al.\ (1991) derived
$\Mloss/\left(d/\kpc\right)^{2}\approx2.0\times10^{-6}
\msun\yr^{-1}$,
corresponding to a mass loss rate of
$\Mloss\approx 1.15\times10^{-4}\msun\yr^{-1}$
for our adopted $d=7.6\kpc$.
Hrivnak et al.\ (2000) modeled
the 2--45$\mum$ ISO spectrum of this source
using DUSTCD.
Assuming $\gastodust = 330$
they derived
$\Mloss/\left\{\left(v_{\rm exp}/\km\s^{-1}\right)\times
\left(d/\kpc\right)\right\}\approx3.2\times10^{-7}
\msun\yr^{-1}$,
corresponding to a mass loss rate of
$\Mloss\approx 3.3\times10^{-5}\msun\yr^{-1}$
for our adopted $d=7.6\kpc$, 
$\vexp=16.0\km\s^{-1}$, 
and $\gastodust = 280$.
Hrivnak \& Bieging (2005) estimated
$\Mloss$\,$\approx$\,$1.4\times10^{-4}\msun\yr^{-1}$
for $d=7.7\kpc$ 
from the CO $J=2-1$ emission line
observed with {\it HHT},
corresponding to
$\Mloss$\,$\approx 1.36\times10^{-4}\msun\yr^{-1}$
for our adopted $d=7.6\kpc$.  

{\it IRAS 23304+6147}:
We derive a mass loss rate of
$\Mloss$\,$\approx$\,$2.39\times10^{-5}\msun\yr^{-1}$
in the AGB phase.
Assuming $\cotoHH = 1\times10^{-3}$,
Likkel et al.\ (1991) derived
$\Mloss/\left(d/\kpc\right)^{2}\approx1.2\times10^{-6}
\msun\yr^{-1}$,
corresponding to a mass loss rate of
$\Mloss\approx 1.33\times10^{-5}\msun\yr^{-1}$
for our adopted $d=3.25\kpc$.
Assuming $\cotoHH = 1\times10^{-3}$,
Omont et al.\ (1993) estimated
$\Mloss$\,$\approx$\,$1.8\times10^{-5}\msun\yr^{-1}$
at $d=2.4\kpc$ from the CO $J=2-1$ line. 
This becomes
$\Mloss\approx 3.3\times10^{-5}\msun\yr^{-1}$
for our adopted $d=3.25\kpc$.
Bakker et al.\ (1997) derived
$\Mloss$\,$\approx$\,$1.0\times10^{-5}\msun\yr^{-1}$
from C$_2$
and
$\Mloss$\,$\approx$\,$5.0\times10^{-5}\msun\yr^{-1}$
from CN (see Footnote~\ref{ftnt:C2CN}).
Hrivnak \& Bieging (2005) estimated
$\Mloss$\,$\approx$\,$2.9\times10^{-4}\msun\yr^{-1}$
for $d=4.7\kpc$ 
from the CO $J=2-1$ emission line
observed with {\it HHT},
corresponding to
$\Mloss$\,$\approx 1.4\times10^{-5}\msun\yr^{-1}$
for our adopted $d=3.25\kpc$. 
%


%

\begin{landscape}
\begin{table*}
\caption{\footnotesize
         \label{tab:starpara}
         Stellar Parameters and {\it IRAS}, 
         {\it WISE}, {\it AKARI}, {\it MSX}, 
         and {\it IRAM}/MAMBO 1.2\,mm Photometry 
         for 18 Galactic 21$\mum$ Sources.
         }
\small\addtolength{\tabcolsep}{-5pt}
\begin{center}
\tiny
\begin{tabular}{lccccccccccccccccccccccccr}
\hline
\hline
IRAS & $\Teff$ & $L_{\star}$ & $M_{\star}$ 
     & $r_{\star}$ & $d$ &  &  
     & {\it IRAS} (Jy)  &  &  &
     & {\it WISE} (Jy) &  & &
     & {\it AKARI} (Jy) &  & &
     & {\it MSX} (Jy) &  &   
     & {\it IRAM} (mJy)\\ 
\cline{7-10}  
\cline{12-15}
\cline{17-18}
\cline{20-23}
        &    &  & & &
        & 12 & 25 & 60 & 100 &  
        & 3.4 & 4.6 & 12 & 22 & 
	& 9& 18 &
	& 8.28 & 12.13 & 14.65 & 21.34
	& 1.2 \\
Sources & (K) & ($L_{\odot}$) & ($M_{\odot}$) 
        & ($R_{\odot}$) & (kpc)
        & ($\mu$m) & ($\mu$m) & ($\mu$m) & ($\mu$m) & 
        & ($\mu$m) & ($\mu$m) & ($\mu$m) & ($\mu$m) &
	& ($\mu$m) & ($\mu$m) &
	& ($\mu$m) & ($\mu$m) & ($\mu$m) & ($\mu$m) 
	& (mm) \\

\hline
Z02229+6208 & 5500  & 8470  & 0.558   & 98.4  & 2.20  & 66.7  & 203.6  & 3.6   & ...   && 2.88  & 4.27 & 42.54 & 168.48 && 29.05 & 110.80 && ...  & ...   & ...  & ...   &... \\
04296+3429  & 6500  & 8334  & 0.554   & 69.1  & 5.00  & 12.7  & 45.9   & 15.5  & 9.22  && 0.30  & 0.47 & 4.85  & 36.25  && 5.99  & 22.33  && ...  & ...   & ...  & ...   &4.4$\pm$1.3 \\
05113+1347  & 5500  & 8315  & 0.604   & 97.5  & 7.00  & 3.78  & 15.3   & 5.5   & 1.67  && 0.32  & 0.28 & 2.45  & 10.68  && 1.48  & 7.60   && ...  & ...   & ...  & ...   &...\\
05341+0852  & 6500  & 8430  & 0.551   & 69.5  & 7.80  & 4.51  & 9.8    & 3.9   & 8.01  && 0.13  & 0.20 & 3.39  & 7.21   && 2.41  & 5.11   && ...  & ...   & ...  & ...   &...\\
06530-0213  & 7000  & 8317  & 0.560   & 59.3  & 4.70  & 6.11  & 27.4   & 15.1  & 4.10  && 0.17  & 0.15 & 6.73  & 21.70  && 2.37  & 13.84  && 1.91 & 7.66  & 9.13 & 18.49 &4.8$\pm$1.5\\
07134+1005  & 7250  & 6555  & 0.841   & 49.0  & 2.40  & 24.5  & 116.7  & 50.1  & 18.7  && 0.84  & 0.61 & 25.74 & 104.46 && 8.91  & 66.25  && ...  & ...   & ...  & ...   &14.0$\pm$1.5\\
07430+1115  & 5500  & 8318  & $<$0.550& 97.5  & 6.00  & 7.68  & 29.9   & 10.7  & 2.53  && 0.32  & 0.34 & 7.43  & 20.71  && 3.30  & 13.50  && ...  & ...   & ...  & ...   &...\\
13245-5036  & 8500  & 2757  & ...     & 23.0  & 7.70  & 1.14  & 3.5    & 1.2   & 1.30  && 0.02  & 0.02 & 0.85  & 3.11   && 0.48  & 2.15   && ...  & ...   & ...  & ...   &...\\
14429-4539  & 6500  & 8552  & ...     & 70.0  & 5.90  & 14.6  & 33.3   & 13.6  & 2.91  && 0.45  & 1.43 & 13.37 & 26.65  && 9.36  & 18.43  && ...  & ...   & ...  & ...   &...\\
15482-5741  & 6500  & 3534  & ...     & 45.0  & 7.00  & 0.67  & 7.1    & 4.0   & 53.80 && ...   & ...  & ...   & ...    && 0.27  & 2.86   && 0.22 & 1.06  & 1.07 & 4.83  &...\\
16594-4656  & 10000 & 10279 & 0.600   & 32.0  & 2.40  & 44.9  & 298.0  & 131.4 & 34.40 && 0.52  & 0.86 & 32.43 & 137.58 && 19.00 & 158    && 15.08& 49.03 & 84.83& 214.07&...\\
19477+2401  & 5500  & 8061  & 0.600   & 96.0  & 5.50  & 11.2  & 54.9   & 27.1  & 38.00 && ...   & ...  & ...   & ...    && 3.46  & ...    && 3.23 & 13.34 & 19.09& 34.3  &...\\
19500-1709  & 8000  & 25435 & 0.599   & 79.0  & 4.10  & 27.8  & 165.0  & 73.4  & 18.20 && 0.66  & 0.55 & 25.71 & 138.98 && ...   & 81.07  && ...  & ...   & ...  & ...   & 29.5\\
20000+3239  & 5500  & 5186  & ...     & 77.0  & 2.24  & 16.0  & 73.6   & 30.0  & 43.10 && 1.34  & 1.73 & 15.78 & 52.64  && 6.78  & 34.03  && 6.15 & 16.53 & 21.37& 44.24 &11.4$\pm$1.7\\
22223+4327  & 6500  & 6075  & 0.551   & 59.0  & 3.20  & 2.12  & 37.1   & 22.4  & 9.54  && 0.52  & 0.36 & 1.72  & 24.69  && 0.79  & 12.23  && ...  & ...   & ...  & ...   &...\\
22272+5435  & 5650  & 10990 & 0.574   & 106.0 & 1.65  & 73.9  & 302.4  & 96.6  & 41.00 && 4.83  & 5.10 & 45.93 & 145.66 && 31.00 & 148.80 && 25.07& 87.85 & 95.38& 186.64&35.3$\pm$1.7\\
22574+6609  & 5500  & 8062  & 0.600   & 96.0  & 7.60  & 9.00  & 29.5   & 20.6  & 7.19  && ...   & ...  & ...   & ...    && 4.73  & 14.80  && ...  & ...   & ...  & ...   &...\\
23304+6147  & 6750  & 8348  & 0.660   & 64.0  & 3.25  & 11.4  & 59.1   & 26.6  & 30.9  && 0.45  & 0.38 & 10.79 & 48.10  && 4.47  & 27.10  && 3.93 & 13.42 & 14.67& 40.96 &...\\
\hline
\end{tabular}
\end{center}
\end{table*}
\end{landscape}

\begin{table*}
\begin{center}
\footnotesize{
\caption{\footnotesize
         \label{tab:drudefit}
          Integrated Fluxes Emitted
          in the UIR ($\Fuir$), 21$\mum$ ($\Ftwenty$), 
          and 30$\mum$ ($\Fthirty$) Features
          as Derived from the PAHFIT Decomposition Method.
          }
\begin{tabular}{lccccccr}
\hline
\hline
IRAS & Warm Dust & Cold Dust & UIR & F$_{21}$ & F$_{30}$ \\    
Sources  & $T_W$ (K) &  $T_C$ (K) & ($10^{-12}\W\m^{-2}$) 
&  ($10^{-12}\W\m^{-2}$) &  ($10^{-12}\W\m^{-2}$) \\
         
\hline
05113+1347 & 150 & 70 & 0.180$\pm$0.002  & 0.047$\pm$0.002 & 0.604$\pm$0.006\\
06530-0213 & 170 & 75 & 0.027$\pm$0.0006 & 0.177$\pm$0.005 & 0.558$\pm$0.004\\
07430+1115 & 180 & 70 & 0.218$\pm$0.003 & 0.035$\pm$0.003 & 1.236$\pm$0.007\\
13245-5036 & 196 & 75 & ... & 0.040$\pm$0.007 & 0.031$\pm$0.006\\
14429-4539\footnotemark[1] & 185 & 76 & 1.194$\pm$0.01 & 0.099$\pm$0.008 & 0.810$\pm$0.021\\
15482-5741 & 130 & 72 & 0.046$\pm$0.005 & 0.040$\pm$0.004 & 0.370$\pm$0.023\\
19477+2401 & ... & 80 & 0.007$\pm$0.0001 & 0.028$\pm$0.001 & 0.521$\pm$0.005\\
19500-1709 & 140 & 90 & 1.034$\pm$0.101 & 0.266$\pm$0.002 & 6.160$\pm$0.129\\
\hline
\end{tabular}
}
\end{center}
\footnotemark[1] A hot component of $T = 445\K$ 
                 is added to fit the continuum
                 at $\lambda < 10\mum$.
\end{table*}

\begin{table*}
\begin{center}
\footnotesize{
\caption{\footnotesize
         \label{tab:modpara}
         Model Parameters for Fitting the Observed SEDs
         Using the 2-DUST Code.
              }
\tiny
\begin{tabular}{lcccccccccccccr}
\hline
\hline
IRAS & $r_{\rm min}$ & $r_{\rm sw}$ & $r_{\rm max}$ & $\tau_{9.8}^{\dag}$ & $\theta^{\ddag}$ & $v_{\rm exp}$
     & $A$ & $B$ & $C$ & $D$ & $E$ & $F$ & $\beta$\\
Sources  & (10$^{16}\cm$) & (10$^{16}\cm$) & (10$^{17}\cm$)
         & &  & (km\,s$^{-1}$) &  & \\
\hline
Z02229+6208  & 2.62$\pm$0.93$^{1}$  & 3.44$\pm$1.28$^{2}$  & 1.71$\pm$0.41        & 0.156$\pm$0.0490 & $90^{\circ a}$ & 14.8$^{5}$ & 8.0   & 6.0  & 1.0  & 5.0  & 9.0  & 10.0  & 5.8 \\
04296+3429   & 4.47$\pm$1.42$^{2}$  & 13.0$\pm$4.18$^{2}$  & 1.43$\pm$0.56        & 0.021$\pm$0.0064 & $45^{\circ b}$ &12.0$^{6}$  & 2.0   & 4.0  & 3.0  & 6.0  & 7.0  & 15.0  & 4.1 \\
05113+1347   & 4.28$\pm$1.21        & 6.10$\pm$2.14        & 1.71$\pm$0.44        & 0.010$\pm$0.0031 & $0^{\circ c}$  &13.1$^{9}$  & 0.1   & 4.5  & 1.0  & 9.0  & 9.0  & 9.0   & 3.5 \\
05341+0852   & 4.10$\pm$1.36        & 4.88$\pm$1.46        & 2.48$\pm$0.68        & 0.002$\pm$0.0005 & $80^{\circ}$   &13.0$^{5}$  & 0.01  & 6.0  & 9.0  & 3.0  & 9.0  & 9.0   & 3.9 \\
06530-0213   & 4.62$\pm$1.59        & 6.01$\pm$1.99        & 2.77$\pm$0.99        & 0.004$\pm$0.0012 & $90^{\circ a}$ &14.0$^{10}$ & 0.01  & 6.0  & 9.0  & 9.0  & 7.0  & 12.0  & 4.0 \\
07134+1005   & 4.29$\pm$1.39$^{3}$  & 6.44$\pm$2.20$^{2}$  & 1.07$\pm$0.41$^{3}$  & 0.027$\pm$0.0091 & $80^{\circ e}$ &10.0$^{6}$  & 3.5   & 3.5  & 1.0  & 7.0  & 9.0  & 3.0   & 4.3 \\
07430+1115   & 2.95$\pm$1.03        & 3.24$\pm$1.00        & 1.50$\pm$0.51        & 0.035$\pm$0.0089 & $90^{\circ c}$ &15.2$^{9}$  & 0.05  & 7.0  & 2.0  & 7.0  & 9.0  & 20.0  & 5.4 \\
13245-5036   & 2.52$\pm$1.01        & 3.27$\pm$1.46        & 0.51$\pm$0.24        & 0.016$\pm$0.0042 & $90^{\circ}$   &15.0        & 5.0   & 3.0  & 3.0  & 6.0  & 4.0  & 8.0   & 3.2 \\
14429-4539   & 1.76$\pm$0.62        & 4.22$\pm$1.77        & 1.77$\pm$0.78        & 0.200$\pm$0.0551 & $90^{\circ}$   &18.2$^{12}$ & 4.0   & 4.0  & 1.0  & 6.0  & 6.0  & 6.0   & 6.8 \\
15482-5741   & 3.23$\pm$1.21        & 4.52$\pm$1.83        & 2.42$\pm$1.01        & 0.076$\pm$0.0258 & $90^{\circ}$   &15.0        & 9.0   & 4.0  & 1.0  & 1.0  & 4.0  & 26.0  & 6.2 \\
16594-4656   & 5.01$\pm$1.38$^{2}$  & 6.01$\pm$1.76$^{2}$  & 2.80$\pm$0.92        & 0.134$\pm$0.0406 & $75^{\circ a}$ &16.0$^{7}$  & 9.0   & 4.5  & 1.0  & 9.0  & 3.0  & 12.0  & 5.3 \\
19477+2401   & 2.62$\pm$0.69        & 4.20$\pm$1.41        & 2.62$\pm$0.81        & 0.385$\pm$0.1033 & $60^{\circ d}$ &13.0$^{5}$  & 5.0   & 6.5  & 3.0  & 6.0  & 9.0  & 3.0   & 4.5 \\
19500-1709   & 10.5$\pm$3.64        & 11.5$\pm$3.87        & 4.72$\pm$1.63        & 0.030$\pm$0.0103 & $45^{\circ b}$ &11.0$^{11}$ & 8.0   & 3.5  & 1.0  & 1.0  & 4.0  & 9.0   & 4.0 \\
20000+3239   & 3.34$\pm$0.87$^{4}$  & 4.01$\pm$1.36        & 1.67$\pm$0.56$^{4}$  & 0.076$\pm$0.0216 & $90^{\circ}$   &12.0$^{6}$  & 12.0  & 4.5  & 1.0  & 3.0  & 1.0  & 12.0  & 5.3 \\
22223+4327   & 7.15$\pm$2.21$^{4}$  & 7.87$\pm$2.61        & 3.58$\pm$1.33$^{4}$  & 0.055$\pm$0.0144 & $90^{\circ f}$ &14.0$^{6}$  & 12.0  & 8.0  & 1.0  & 4.0  & 3.0  & 15.0  & 5.4 \\
22272+5435   & 4.62$\pm$0.88        & 5.10$\pm$1.53        & 1.88$\pm$0.46        & 0.067$\pm$0.0221 & $45^{\circ b}$ &9.80$^{5}$  & 12.0  & 5.5  & 1.0  & 1.0  & 1.0  & 15.0  & 4.9 \\
22574+6609   & 2.04$\pm$0.58$^{5}$  & 4.89$\pm$1.48        & 1.20$\pm$0.41        & 0.361$\pm$0.1176 & $90^{\circ d}$ &16.0$^{8}$  & 2.0   & 4.0  & 3.0  & 4.0  & 5.0  & 9.0   & 4.8 \\
23304+6147   & 5.33$\pm$1.44$^{4}$  & 6.92$\pm$2.03        & 2.66$\pm$0.87$^{4}$  & 0.039$\pm$0.0128 & $90^{\circ g}$ &15.5$^{6}$  & 12.0  & 6.0  & 5.0  & 3.5  & 7.0  & 12.0  & 5.5 \\
\hline
\end{tabular}
}
\end{center}
$^{\dag}$ Optical depth at 9.8$\mum$.\\
$^{\ddag}$ Inclination angle:
$^{a}$ Ueta et al.\ (2005);
$^{b}$ Meixner et al.\ (1997);
$^{c}$ Ueta et al.\ (2000);
$^{d}$ Su et al.\ (2001);
$^{e}$ Meixner et al.\ (2004);
$^{f}$ Hrivnak et al.\ (2011);
$^{g}$ Bujarrabal et al.\ (2001).
\\
$^{1}$ Kwok et al.\ (2002);
$^{2}$ Ueta et al.\ (2005);
$^{3}$ Nakashima et al.\ (2009);
$^{4}$ Sahai et al.\ (2007);
$^{5}$ Hrivnak et al.\ (2000);
$^{6}$ Bakker et al.\ (1997);
$^{7}$ Loup et al.\ (1990);
$^{8}$ Hrivnak \& Kwok (1991);
$^{9}$ Hrivnak \& Kwok (1999);
$^{10}$ Hrivnak \& Reddy (2003);
$^{11}$ Likkel et al.\ (1987);
$^{12}$ Reddy \& Parthasarathy (1996).
%
\end{table*}

\begin{table*}
\begin{center}
\footnotesize{
\caption{\footnotesize
         \label{tab:Feat2Photo}
         Contributions of the 21 and 30$\mum$ Features
         to the Photometric Fluxes of 
         the {\it AKARI} 18$\mum$ Band,
         the {\it MSX} $E$ Band at 21.34$\mum$,
         the {\it WISE} 22$\mum$ Band,
         and the {\it IRAS} 25$\mum$ Band.
         }
         
\begin{tabular}{lcccccr}
\hline
\hline
IRAS     & {\it IRAS} (25\,$\mu$m)      
         & {\it WISE} (22\,$\mu$m)      
         & {\it AKARI} (18\,$\mu$m)       
         & {\it MSX} (21.34\,$\mu$m)\\ 
Sources  & (Jy) & (Jy) & (Jy) &  (Jy) \\
\hline
Z02229+6208   & 37.89 & 2.82  & 13.04 & 13.67  \\
04296+3429    & 9.21  & 1.12  & 4.75   & 5.34     \\
05113+1347    & 2.41  & 0.24  & 1.05   & 1.14     \\
05341+0852    & 1.33  & 0.12  & 0.54   & 0.58     \\
06530-0213    & 3.68   & 0.55  & 2.27   & 2.59     \\
07134+1005    & 18.97 & 3.00  & 12.44 & 14.33   \\
07430+1115    & 4.64   & 0.37  & 1.66   & 1.75      \\
13245-5036    & 0.63   & 0.10  & 0.42   & 0.46      \\
14429-4539    & 3.65   & 0.40  & 1.69   & 1.88      \\
15482-5741    & 1.39   & 0.15  & 0.65   & 0.72      \\
16594-4656    & 60.75 & 7.56  & 33.79 & 37.00   \\
19477+2401    & 2.42   & 9.26  & 1.21   & 1.35      \\
19500-1709    & 16.81 & 1.21  & 6.63   & 5.27      \\
20000+3239    & 9.03   & 0.90  & 4.04   & 4.30       \\
22223+4327    & 4.72   & 0.49  & 2.11   & 2.33       \\
22272+5435    & 48.28 & 4.39  & 19.40 & 21.01    \\
22574+6609    & 4.08   & 0.42  & 2.28   & 2.33       \\
23304+6147    & 11.95 & 1.32  & 5.97   & 6.62       \\
\hline
\end{tabular}
}
\end{center}
\end{table*}

\begin{table*}
\footnotesize{
\caption{\footnotesize
         \label{tab:massloss}
         Dust Mass ($\Md$) 
         and Stellar Mass Loss Rates ($\dot{M}$) 
         in the AGB and SW Phases 
         (Assuming a Gas-to-Dust Ratio of $\sim$\,280 
          Appropriate for Carbon-Rich AGB Stars;
          Justtanont et al.\ 1996)
          as Derived from the 2-DUST Model
          of Ueta \& Meixner (2003).
         }
\begin{center}
\tiny
\begin{tabular}{lcccccccccr}
\hline
\hline
IRAS  & &  & AGB &  &  &  & SW & \\ 
      \cline{3-5} \cline{7-9} Sources & 
      &  &  & & &  &  &  \\	
      &  & $\dot{M}$ & $\Md$ & Duration &  
      & $\dot{M}$ & $\Md$ & Duration \\
      &  & ($M_{\odot}\yr^{-1})$ & ($M_{\odot}$) 
      & (yr) &  &  ($M_{\odot}\yr^{-1}$) 
      & ($M_{\odot}$) & (yr)  \\
\hline
Z02229+6208&&(1.94$\pm$0.68)$\times 10^{-5}$&(1.96$\pm$0.54)$\times 10^{-4}$&(2.83$\pm$0.89)$\times 10^{3}$&&(5.53$\pm$2.36)$\times 10^{-4}$&(3.79$\pm$1.13)$\times 10^{-4}$&(1.92$\pm$0.50)$\times 10^{2}$\\
04296+3429&& (3.84$\pm$1.32)$\times 10^{-5}$&(4.88$\pm$1.55)$\times 10^{-5}$&(3.55$\pm$0.92)$\times 10^{2}$&&(1.29$\pm$0.58)$\times 10^{-4}$&(1.04$\pm$0.35)$\times 10^{-3}$&(2.25$\pm$0.67)$\times 10^{3}$\\
05113+1347&& (2.59$\pm$1.02)$\times 10^{-5}$&(3.40$\pm$1.01)$\times 10^{-4}$&(3.67$\pm$0.86)$\times 10^{3}$&&(1.24$\pm$0.49)$\times 10^{-4}$&(1.81$\pm$0.55)$\times 10^{-4}$&(4.08$\pm$1.25)$\times 10^{2}$\\
05341+0852&& (1.01$\pm$0.37)$\times 10^{-5}$&(1.63$\pm$0.42)$\times 10^{-4}$&(4.53$\pm$1.09)$\times 10^{3}$&&(1.77$\pm$0.81)$\times 10^{-4}$&(1.17$\pm$0.38)$\times 10^{-4}$&(1.85$\pm$0.59)$\times 10^{1}$\\
06530-0213&& (1.86$\pm$0.66)$\times 10^{-5}$&(3.06$\pm$0.86)$\times 10^{-4}$&(4.61$\pm$1.01)$\times 10^{3}$&&(2.36$\pm$0.92)$\times 10^{-4}$&(2.48$\pm$0.76)$\times 10^{-4}$&(2.94$\pm$0.62)$\times 10^{2}$\\
07134+1005&& (1.99$\pm$0.83)$\times 10^{-5}$&(2.33$\pm$0.62)$\times 10^{-4}$&(1.36$\pm$0.31)$\times 10^{3}$&&(1.24$\pm$0.50)$\times 10^{-4}$&(3.02$\pm$0.95)$\times 10^{-4}$&(6.82$\pm$1.75)$\times 10^{2}$\\
07430+1115&& (5.78$\pm$2.45)$\times 10^{-5}$&(5.17$\pm$1.53)$\times 10^{-4}$&(2.50$\pm$0.62)$\times 10^{3}$&&(1.05$\pm$0.38)$\times 10^{-3}$&(2.35$\pm$0.81)$\times 10^{-4}$&(6.26$\pm$1.50)$\times 10^{1}$\\
13245-5036&& (4.98$\pm$2.39)$\times 10^{-5}$&(6.66$\pm$2.36)$\times 10^{-5}$&(3.75$\pm$1.21)$\times 10^{2}$&&(7.36$\pm$3.45)$\times 10^{-5}$&(4.22$\pm$1.46)$\times 10^{-5}$&(1.61$\pm$0.51)$\times 10^{2}$\\
14429-4539&& (3.22$\pm$1.63)$\times 10^{-5}$&(3.06$\pm$1.22)$\times 10^{-4}$&(2.65$\pm$0.83)$\times 10^{3}$&&(2.97$\pm$1.42)$\times 10^{-4}$&(5.18$\pm$1.91)$\times 10^{-4}$&(4.89$\pm$1.49)$\times 10^{2}$\\
15482-5741&& (2.50$\pm$1.31)$\times 10^{-5}$&(3.73$\pm$1.51)$\times 10^{-4}$&(4.18$\pm$1.41)$\times 10^{3}$&&(2.37$\pm$1.19)$\times 10^{-4}$&(2.32$\pm$0.91)$\times 10^{-4}$&(2.74$\pm$0.86)$\times 10^{2}$\\
16594-4656&& (8.17$\pm$2.62)$\times 10^{-5}$&(1.46$\pm$0.51)$\times 10^{-3}$&(5.01$\pm$1.23)$\times 10^{3}$&&(7.98$\pm$3.93)$\times 10^{-4}$&(6.48$\pm$1.78)$\times 10^{-4}$&(2.27$\pm$0.62)$\times 10^{2}$\\
19477+2401&& (2.37$\pm$1.09)$\times 10^{-5}$&(4.57$\pm$1.21)$\times 10^{-4}$&(5.39$\pm$1.33)$\times 10^{3}$&&(1.44$\pm$0.49)$\times 10^{-3}$&(1.99$\pm$0.66)$\times 10^{-3}$&(3.85$\pm$0.96)$\times 10^{2}$\\
19500-1709&& (7.14$\pm$3.14)$\times 10^{-5}$&(2.64$\pm$0.76)$\times 10^{-3}$&(1.03$\pm$0.26)$\times 10^{4}$&&(4.34$\pm$1.54)$\times 10^{-4}$&(4.72$\pm$1.61)$\times 10^{-4}$&(3.04$\pm$0.83)$\times 10^{2}$\\
20000+3239&& (2.32$\pm$1.02)$\times 10^{-5}$&(2.79$\pm$0.81)$\times 10^{-4}$&(3.36$\pm$0.87)$\times 10^{3}$&&(1.95$\pm$0.79)$\times 10^{-4}$&(1.23$\pm$0.44)$\times 10^{-4}$&(1.77$\pm$0.59)$\times 10^{2}$\\
22223+4327&& (2.18$\pm$0.90)$\times 10^{-5}$&(4.93$\pm$1.36)$\times 10^{-4}$&(6.34$\pm$1.90)$\times 10^{3}$&&(7.10$\pm$3.21)$\times 10^{-4}$&(4.12$\pm$1.57)$\times 10^{-4}$&(1.62$\pm$0.43)$\times 10^{2}$\\
22272+5435&& (2.59$\pm$1.22)$\times 10^{-5}$&(3.75$\pm$1.23)$\times 10^{-4}$&(4.05$\pm$1.13)$\times 10^{3}$&&(3.50$\pm$1.38)$\times 10^{-4}$&(1.70$\pm$0.56)$\times 10^{-4}$&(1.36$\pm$0.41)$\times 10^{2}$\\
22574+6609&& (3.39$\pm$1.21)$\times 10^{-4}$&(1.72$\pm$0.50)$\times 10^{-3}$&(1.42$\pm$0.35)$\times 10^{3}$&&(1.39$\pm$0.56)$\times 10^{-3}$&(2.81$\pm$0.91)$\times 10^{-3}$&(5.67$\pm$1.57)$\times 10^{2}$\\
23304+6147&& (2.39$\pm$1.07)$\times 10^{-5}$&(3.45$\pm$1.21)$\times 10^{-4}$&(4.05$\pm$1.12)$\times 10^{3}$&&(4.17$\pm$1.52)$\times 10^{-4}$&(4.87$\pm$1.51)$\times 10^{-4}$&(3.28$\pm$0.63)$\times 10^{2}$\\

\hline
\end{tabular}
\end{center}
}
\end{table*}

\begin{deluxetable}{lcccccr}
\tablecolumns{5}
\tablewidth{0pt}
\center
\tablecaption{\footnotesize
              \label{tab:multivariate}
              Correlation Matrix among
              the 21 and 30$\mum$
              Features and the AGB and SW
              Mass Loss Rates
              Derived from the PCA
              Multivariate Analysis
              \vspace{5mm}
              }
\tablehead{
\colhead{}&
\colhead{$\Ftwenty$}&
\colhead{$\Fthirty$}&
\colhead{$\dMAGB/d^{2}$}&
\colhead{$\dMSW/d^{2}$}&
}
\startdata
$\Ftwenty$ & 1.00 & 0.64 & 0.82  & 0.62 \\
$\Fthirty$ & 0.64 & 1.00 & 0.87 & 0.83  \\
$\dMAGB/d^{2}$ & 0.82  & 0.87 & 1.00 & 0.80 \\
$\dMSW/d^{2}$  & 0.62 & 0.83  & 0.80 & 1.00\\
\enddata  
\end{deluxetable}

%

\begin{deluxetable}{lccc}
\tabletypesize{\tiny}
\tablewidth{380pt}
\tablecaption{\footnotesize
         \label{tab:massloss}
         Comparison of the Mass Loss Rates
         ($\Mloss$) Derived in This Work
         with That Reported in the Literature.
         All Are Based on the Same Distance
         and Gas-to-Dust Ratio as Those Adopted 
         in This Work.
         }
\tablehead{\colhead{IRAS}     
           & \colhead{$\dot{M}$}         
           & \colhead{References} 
           & \colhead{Methodology} 
\\
\colhead{Sources}  & \colhead{($\msun\yr^{-1}$)} &  & 
}
\startdata
Z02229+6208 & AGB: 1.94$\times 10^{-5}$, SW: 5.53$\times 10^{-4}$ 
            & This work                  & IR SED \\
            & 1.4$\times 10^{-4}$        
            & Hrivnak \& Bieging (2005)  & CO \\
            & 3.59$\times 10^{-5}$       
            & Hrivnak et al.\ (2000)     & IR SED \\
\\       
04296+3429  & AGB: 3.84$\times 10^{-5}$, SW: 1.29$\times 10^{-4}$ 
            & This work                  & IR SED \\
            & 3.10$\times 10^{-5}$                    
            & Hrivnak \& Bieging (2005)  & CO   \\
            & 1.75$\times 10^{-5}$                   
            & Sahai (1999)                & Scattered Light \\
            & 6.30$\times 10^{-6}$                    
            & Bakker et al.\ (1997)       & CN \\
            & 1.60$\times 10^{-6}$                    
            & Bakker et al.\ (1997)        & C$_{2}$ \\
            & 4.00$\times 10^{-5}$                    
            & Bujarrabal et al.\ (2001)   & CO   \\
            & 3.10$\times 10^{-5}$                    
            & Meixner et al.\ (1997)       & IR SED \\
\\
05113+1347  & AGB: 2.59$\times 10^{-5}$, SW: 1.24$\times 10^{-4}$  
            & This work                  & IR SED \\
            & 7.90$\times 10^{-6}$                    
            & Bakker et al.\ (1997)        & CN \\
            & 7.90$\times 10^{-7}$                    
            & Bakker et al.\ (1997)        & C$_{2}$ \\
            & 6.22$\times 10^{-4}$                    
            & Reddy \& Parthasarathy (1996) & IRAS 60$\mum$ Emission \\
            & 2.7$\times 10^{-4}$                   
            & Hrivnak et al.\ (2009)       & IR SED\\
\\          
05341+0852  & AGB: 1.01$\times 10^{-5}$, SW: 1.77$\times 10^{-4}$ 
            & This work                  & IR SED \\
            & 1.00$\times 10^{-5}$                    
            & Bakker et al.\ (1997)        & CN \\
            & 1.00$\times 10^{-6}$                    
            & Bakker et al.\ (1997)        & C$_{2}$ \\  
            & 1.60$\times 10^{-7}$                    
            & Reddy \& Parthasarathy (1996) & IRAS 60$\mum$ Emission \\  
            & 7.68$\times 10^{-5}$                    
            & Hrivnak et al.\ (2009)       & IR SED \\
\\
06530-0213  & AGB: 1.86$\times 10^{-5}$, SW: 2.36$\times 10^{-4}$ 
            & This work                  & IR SED \\
            & 4.42$\times 10^{-7}$                    
            & Reddy \& Parthasarathy (1996) & IRAS 60$\mum$ Emission \\  
            & 3.30$\times 10^{-5}$                    
            & Hrivnak \& Bieging (2005)  & CO  \\
            & 7.71$\times 10^{-5}$                    
            & Hrivnak et al.\ (2009)       & IR SED \\
\\
07134+1005  & AGB: 1.99$\times 10^{-5}$, SW: 1.24$\times 10^{-4}$ 
            & This work                  & IR SED \\
            & 9.7$\times 10^{-6}$                    
            & Omont et al.\ (1993)         & CO  \\
            & 1.2$\times 10^{-4}$                    
            & Hony et al.\ (2003)          & IR SED\\
            & 5.1$\times 10^{-6}$                    
            & Meixner et al.\ (2004)       & CO  \\
            & 2.7$\times 10^{-5}$                    
            & Meixner et al.\ (2004)       & IR SED \\
            & 2.3$\times 10^{-5}$                    
            & Hrivnak \& Bieging (2005)  & CO  \\
            & 4.43$\times 10^{-5}$                   
            & Buemi et al.\ (2007)         & IR SED \\
            & 1.47$\times 10^{-5}$                   
            & Hrivnak et al.\ (2000)       & IR SED \\
\\
07430+1115  & AGB: 5.78$\times 10^{-5}$, SW: 1.05$\times 10^{-3}$ 
            & This work                  & IR SED \\
            & 7.48$\times 10^{-4}$                    
            & Hrivnak et al.\ (2009)       & IR SED\\
\\
14429-4539  & AGB: 3.22$\times 10^{-5}$, SW: 2.97$\times 10^{-4}$ 
            & This work                 & IR SED \\
            & 3.8$\times 10^{-7}$                    
            & Reddy \& Parthasarathy (1996) & IRAS 60$\mum$ Emission \\  
\\        
16594-4656 & AGB: 8.17$\times 10^{-5}$, SW: 7.98$\times 10^{-4}$ 
           & This work                  & IR SED \\
           & 1.27$\times 10^{-4}$                   
           & Hrivnak et al.\ (2000)       & IR SED \\
\\           
19477+2401 & AGB: 2.37$\times 10^{-5}$, SW: 1.44$\times 10^{-3}$ 
           & This work                  & IR SED \\
           & 1.52$\times 10^{-4}$                   
           & Hrivnak et al.\ (2000)       & IR SED \\
\\   
19500-1709 & AGB: 7.14$\times 10^{-5}$, SW: 4.34$\times 10^{-4}$ 
           & This work                  & IR SED \\
           & 8.97$\times 10^{-5}$                    
           & Clube \& Gledhill (2004)   & IR SED \\
           & 6.89$\times 10^{-5}$                   
           & Meixner et al.\ (1997)       & IR SED\\
           & 2.02$\times 10^{-5}$                   
           & Likkel et al.\ (1991)        & CO  \\
           & 3.05$\times 10^{-5}$                    
           & Omont et al.\ (1993)         & CO \\
\\
20000+3239 & AGB: 2.32$\times 10^{-5}$, SW: 1.95$\times 10^{-3}$ 
           & This work                 & IR SED \\
           & 3.47$\times 10^{-5}$                    
           & Buemi et al.\ (2007)        & IR SED \\
           & 5.52$\times 10^{-6}$                    
           & Likkel et al.\ (1991)       & CO  \\
           & 1.16$\times 10^{-5}$                    
           & Hrivnak et al.\ (2000)      & IR SED \\
           & 1.99$\times 10^{-6}$                     
           & Omont et al.\ (1993)        & CO \\        
\\
22223+4327 & AGB: 2.18$\times 10^{-5}$, SW: 7.10$\times 10^{-4}$ 
           & This work                 & IR SED \\
           & 6.14$\times 10^{-6}$                    
           & Likkel et al.\ (1991)       & CO  \\
           & 1.6$\times 10^{-4}$                     
           & Bakker et al.\ (1997)       & CN \\
           & 2.5$\times 10^{-5}$                     
           & Bakker et al.\ (1997)       & C$_{2}$ \\   
\\       
22272+5435 & AGB: 2.59$\times 10^{-6}$, SW: 3.50$\times 10^{-4}$ 
           & This work                 & IR SED \\
           & 1.88$\times 10^{-5}$                     
           & Bujarrabal et al.\ (2001)   & CO  \\
           & 4.2$\times 10^{-5}$                    
           & Buemi et al.\ (2007)        &IR SED \\
           & 5.8$\times 10^{-5}$                     
           & Hrivnak \& Bieging  (2005) & CO \\
\\
22574+6609 & AGB: 3.39$\times 10^{-5}$, SW: 1.39$\times 10^{-3}$ 
           & This work                 & IR SED \\
           & 1.15$\times 10^{-4}$                    
           & Likkel et al.\ (1991)       & CO \\
           & 1.36$\times 10^{-4}$                     
           & Hrivnak \& Bieging  (2005) & CO \\
           & 3.3$\times 10^{-5}$                    
           & Hrivnak et al.\ (2000)      & IR SED \\
\\      
23304+6147 & AGB: 2.39$\times 10^{-5}$, SW: 4.17$\times 10^{-4}$ 
           & This work                 & IR SED \\
           & 1.33$\times 10^{-5}$                    
           & Likkel et al.\ (1991)       & CO \\
           & 1.4$\times 10^{-5}$                     
           & Hrivnak \& Bieging  (2005) & CO \\
           & 3.35$\times 10^{-5}$                     
           & Omont et al.\ (1993)        & CO \\
           & 5.0$\times 10^{-5}$                     
           & Bakker et al.\ (1997)       & CN \\
           & 1.0$\times 10^{-5}$                     
           & Bakker et al.\ (1997)       & C$_{2}$ \\  
\enddata
\end{deluxetable}


\end{document}